\documentclass[twocolumn]{emulateapj}
\slugcomment{{\sc Accepted to ApJ:} 13 May , 2012}
\usepackage{natbib,amsmath}
\newcommand{\cmt}{{\rm cm}$^{-3}$}

\newcommand{\pcsq}{{\rm pc}$^2$}
\newcommand{\tm}{$\times$}

\newcommand{\msun}{M$_\odot$}
\newcommand{\msunpc}{M$_\odot$ pc$^{-2}$}
\newcommand{\kms}{km s$^{-1}$}

\newcommand{\vz}{$\sigma_{v}$}
\newcommand{\Ht}{$H$}
\newcommand{\sig}{$\Sigma$}
\newcommand{\omg}{$\Omega$}
\newcommand{\sigmol}{$\Sigma_{\rm mol}$}
\newcommand{\sigsfr}{$\Sigma_{\rm SFR}$}
\newcommand{\sfrunits}{${\rm M}_\odot \, {\rm kpc}^{-2} \, {\rm yr}^{-1}$}
\newcommand{\psn}{$p_{*}$}
\newcommand{\rsh}{$R_{\rm sh}$}
\newcommand{\Nz}{$N_{\rm z}$}
\newcommand{\Nr}{$N_{\rm R}$}
\newcommand{\Nsn}{$N_{\rm SN}$}
\newcommand{\Lz}{$L_{\rm z}$}
\newcommand{\Lr}{$L_{\rm R}$}
\newcommand{\Lphi}{$L_{\rm \phi}$}
\newcommand{\msn}{$m_*$}
\newcommand{\nth}{$n_{\rm th}$}

\newcommand{\torb}{$t_{\rm orb}$}
\newcommand{\dtbin}{$\Delta t_{\rm bin}$}

\newcommand{\tffnth}{$t_{\rm ff}(n_{\rm th})$}
\newcommand{\tffno}{$t_{\rm ff}(n_{0})$}

\newcommand{\epsnth}{$\epsilon_{\rm ff}(n_{\rm th})$}
\newcommand{\epsrhoo}{$\epsilon_{\rm ff}(n_{\rm 0})$}

\newcommand{\no}{$n_{\rm 0}$}
\newcommand{\fup}{$f_{p}$}
\newcommand{\pturb}{$P_{\rm turb}$}
\newcommand{\pdriv}{$P_{\rm drive}$}

\newcommand {\apgt} {\ {\raise-.5ex\hbox{$\buildrel>\over\sim$}}\ }
\newcommand {\aplt} {\ {\raise-.5ex\hbox{$\buildrel<\over\sim$}}\ } 
\newcommand{\XCO}{$X_{\rm CO}$}
\shorttitle{Modeling Starburst Regulation}
\shortauthors{Shetty \& Ostriker}

\begin{document}
\title{Maximally Star-Forming Galactic Disks II. Vertically-Resolved Hydrodynamic Simulations of Starburst Regulation}
\author{Rahul Shetty\altaffilmark{1} and Eve C. Ostriker\altaffilmark{2}}
\altaffiltext{1}{Zentrum f\"ur Astronomie der Universit\"at Heidelberg, Institut f\"ur Theoretische Astrophysik, Albert-Ueberle-Str. 2, 69120 Heidelberg, Germany; R.Shetty@.uni-heidelberg.de}
\altaffiltext{2}{Department of Astronomy, University of Maryland, College Park, MD 20742, USA; ostriker@astro.umd.edu}

\begin{abstract}

We explore the self-regulation of star formation using a large suite
of high resolution hydrodynamic simulations, focusing on
molecule-dominated regions (galactic centers and [U]LIRGS) where
feedback from star formation drives highly supersonic turbulence.  In
equilibrium the total midplane pressure, dominated by turbulence, must
balance the vertical weight of the ISM.  Under self-regulation, the
momentum flux injected by feedback evolves until it matches the
vertical weight.  We test this flux balance in simulations spanning a
wide range of parameters, including surface density \sig, momentum
injected per stellar mass formed ($p_*/m_*$), and angular velocity.
The simulations are two dimensional radial-vertical slices, and
include both self-gravity and an external potential that helps to
confine gas to the disk midplane.  After the simulations reach a
steady state in all relevant quantities, including the star formation
rate \sigsfr, there is remarkably good agreement between the vertical
weight, the turbulent pressure, and the momentum injection rate from
supernovae.  Gas velocity dispersions and disk thicknesses increase
with $p_*/m_*$.  The efficiency of star formation per free-fall time
at the mid-plane density, \epsrhoo, is insensitive to the local
conditions and to the star formation prescription in very dense gas.
We measure \epsrhoo$\sim$0.004-0.01, consistent with low and
approximately constant efficiencies inferred from observations.  For
\sig$\in$(100--1000) \msunpc, we find \sigsfr$\in$(0.1--4) \sfrunits,
generally following a \sigsfr$\propto$ \sig$^2$ relationship.  The
measured relationships agree very well with vertical equilibrium and
with turbulent energy replenishment by feedback within a vertical
crossing time.  These results, along with the observed
\sig--\sigsfr\ relation in high density environments, provide strong
evidence for the self-regulation of star formation.

\end{abstract}


\section{Introduction\label{introsec}}

\subsection{Star Formation on Galactic Scales}

Observations reveal that stars form in the molecular component of the
interstellar medium (ISM).  Therefore, the dynamics of molecular gas
must affect the star formation process.  On galactic scales, gravity
concentrates gas into clouds in which stars eventually form.  The
resulting feedback from stellar winds, ionizing and non-ionizing
radiation, and supernovae (SN) (either local or nearby in the disk)
redisperses this dense gas.  The formation, destruction, and the
dynamical state of star forming clouds depend strongly on the local
conditions of the ISM.  In (ultra) luminous infrared galaxies
([U]LIRGs) and the centers of galaxies, molecular gas pervades much of
the ISM, including regions not actively forming stars.  Gas in such
environments has higher mean volume and surface density compared to
the gas found in giant molecular clouds
\citep[GMCs,][]{SolomonVandenBout05} in lower-density regions of
galaxies.  Near-future ALMA observations will resolve high density
tracers, and thereby reveal the detailed structure and kinematics of
gas in starbursts.  Understanding how small-scale feedback associated
with star-formation acts in concert with larger scale processes in
starbursts (as well as mid-- and outer-- disks) is crucial for
developing any successful theory of galactic star formation.

Stellar feedback plays a key role in regulating the thermal balance
and morphological structure of the ISM \citep{McKee&Ostriker77,
  Norman&Ikeuchi89}.  Feedback is also believed to be the primary
mechanism driving turbulence \citep[e.g.][]{Norman&Ferrara96}.  Since
turbulence is observed on all scales larger than the size of the
densest prestellar cores, it is now understood to be an essential
component controlling the dynamics and regulating star formation in
the ISM \citep[see][and references
  therein]{MacLow&Klessen04,McKee&Ostriker07}.  The vertical scale
height of the galactic disk depends on the balance between gaseous,
stellar, and dark matter potentials that concentrate gas, and the
pressures (thermal, turbulent, magnetic, cosmic ray, and radiation)
that oppose gravity and limit runaway collapse
\citep[e.g.][]{Boulares&Cox90}.

Over sufficiently large scales, the star formation rate surface
density, \sigsfr, is observed to correlate well with the gas surface
density, \sig\ \citep{Schmidt59, Kennicutt89,Kennicutt98}.  This
correlation appears to take on various forms in different regions
within disk galaxies.  In the outer-disk regions containing little
molecular gas, there is no universal power-law index describing the
\sigsfr$-$\sig\ relationship \citep{Bigiel+10}.  Instead,
\sigsfr\ depends on both \sig\ and the stellar density
\citep{Blitz&Rosolowsky04,Blitz&Rosolowsky06}; this is presumably
because stellar rather than gas vertical gravity dominates in outer
disks (see below).  At smaller radii, by mass the ISM is dominated by
molecular gas, for which two different star formation laws appear to
take hold.  In mid-disk regions where most of the volume is filled
with atomic gas and molecules are confined in isolated GMCs (with a
limited range of properties -- \citet{Sheth+08,Bolattoetal08}), there
is a strong, approximately linear correlation between the star
formation rate and molecular mass, \sigsfr\ $\propto$
\sigmol\ \citep{Wong&Blitz02, Bigiel+08, Schruba+11}.  Towards the
central regions and in starbursts where the ISM is almost completely
molecular \citep{SolomonVandenBout05}, there appears to be a steeper
\sigsfr -- \sigmol\ relationship \citep[][]{Kennicutt98,Genzel+10,
  Daddi+10, Narayanan+12}.

The variations in \sig--\sigsfr\ correlations in different galactic
regions presumably owe to the differences in the characteristics of
the ISM.  Gas properties such as the temperatures, densities, and
velocities are found to vary between starbursts and more quiescent
environments.  In the Galactic center, molecular gas is much more
prevalent \citep[e.g.][]{Bally+87, Bally+88}, and \sigsfr\ is measured
to be $\sim$1.5 dex higher than in the mid- to outer- disk
\citep{Yusef-Zadeh+09}.  Observed linewidths from the dense, molecular
gas in the Galactic Center are measured to reach \apgt 10
\kms\ \citep{Oka+98, Oka+01, Shetty+12}, and as high as \apgt 100
\kms\ in starbursts \citep{Solomonetal97,Downes&Solomon98, Genzel+11}.
Turbulent velocities in GMCs are significantly lower, ranging from 1
-- 6 \kms\ \citep[e.g.][]{Larson81, Solomonetal87}.  However, present
observations of (U)LIRGs do not have sufficient resolution to
distinguish between perturbed motions (such as large-scale streaming)
on scales $\apgt H$, the disk thickness, and more localized turbulence
(i.e. velocity dispersions on $\sim$10 pc scales, similar to GMCs).

Global numerical simulations of disk galaxies have shown that the
structures formed by self-gravity \citep[e.g.][]{Shetty&Ostriker06,
  Dobbs&Pringle09, Tasker&Tan09, Tasker11} or by cloud collisions
\citep[e.g.][]{Dobbs08, Tasker&Tan09} can generally reproduce observed
morphological features of the ISM, such as filamentary substructure,
cloud masses, sizes, and basic kinematic properties.  Additionally,
large scale simulations have suggested that gravitational instability
naturally results in power-law relationships between \sigsfr\ and
\sig\ if the Toomre $Q$ and velocity dispersion are uniform
\citep[e.g.][]{Lietal05, Lietal06, Shetty&Ostriker08}.  Simulations
with feedback have produced a range in the exponent and coefficient of
the \sigsfr-\sig\ relationship, depending on the specific feedback
prescription \citep[e.g][]{Tasker&Bryan06, Tasker&Bryan08,
  Robertson&Kravtsov08, Shetty&Ostriker08, KoyamaOstriker09a,
  Dobbs+11b, Hopkins+11}.  \citet{Shetty&Ostriker08} pointed out that
the relationship between \sigsfr\ and \sig\ in general should depend
on the thickness of the gas disk, and therefore on the gas velocity
dispersion and on the stellar potential if it dominates (see below).

Variations in feedback parameters, such as the injected momenta,
energies, and rates, combined with other processes such as rotation,
vertical motions due to an external potential, shear, and large-scale
gravitational instability in the shearing, rotating flow, are likely
to contribute to the observed differences in velocity dispersions
between starbursts and more quiescent regions.
\citet{Ostriker&Shetty11} and \citet{Kim+11} (hereafter KKO11) argue
that the velocity dispersion on scales comparable to the neutral gas
disk's thickness will be relatively constant if turbulence is driven
by feedback, because the driving rate and dissipation rate both scale
inversely with the vertical crossing time (or gravitational free-fall
time) of the ISM.  Simulations of the ISM in mid-- and outer-- disk
environments have shown that velocity dispersions are in fact not
strongly sensitive to the feedback parameters
\citep[e.g.][KKO11]{Dib+06, Shetty&Ostriker08, Joung+09}.  Such
simulations allow for a detailed assessment of the relationships
between the relevant physical quantities, and provide a direct avenue
for testing analytical theories of star formation.

\subsection{Theory of Star Formation Self-Regulation}

A theory for the self-regulation of star formation on galactic scales
has recently been formulated by \citet{Ostriker+10} (hereafter OML10)
and \citet{Ostriker&Shetty11} (hereafter Paper I).  KKO11 conducted
numerical models of multi-phase gaseous disks in the regime where
diffuse atomic gas dominates (\sig\ \aplt 20 \msunpc), verifying the
assumptions and predicted features of the self-regulated
thermal/dynamical equilibrium theory.  In the present work, we shall
instead focus on numerical simulations of the molecule-dominated
starburst regime.  To provide an overall context and distinguish
between the various regimes, we briefly review the concepts and
analysis of the self-regulation model.

For dynamical equilibrium to be satisfied, the total pressure at the
midplane must balance the gravitational weight of the overlying
diffuse-ISM gas, $P_{\rm tot} = W \equiv (1/2)\Sigma_{\rm diff}
\langle g_z \rangle$.  In different regimes, this pressure may be
dominated by different terms (thermal, turbulent, or radiation), but
each pressure term individually responds to the star formation rate.
Where there is a substantial amount of atomic gas heated by stellar
UV, balance of heating and cooling leads to an equilibrium thermal
pressure $P_{\rm th}\propto J_{\rm UV} \propto \Sigma_{\rm SFR}$
(OML10, KKO11).  Similarly, balancing turbulent driving associated
with expanding radiative SN remnants (or other massive-star momentum sources)
with dissipation on a vertical crossing time leads to an equilibrium
turbulent pressure $P_{\rm turb} \propto \Sigma_{\rm SFR}$ (Paper I,
KKO11).  In extremely high \sig\ regions, trapped reprocessed
starlight provides a radiation pressure $P_{\rm rad} \propto \Sigma
\Sigma_{\rm SFR}$ that begins to compete with the turbulent pressure
\citep[][Paper I]{Thompson+05}.  Putting these individual terms
together, $P_{\rm tot} = P_{\rm th} + P_{\rm turb} + P_{\rm rad}
\propto \Sigma_{\rm SFR}$.  Thus, under self-regulation the combined
constraints of thermal, turbulent, radiative, and dynamical
equilibrium imply that the star formation rate will naturally evolve
to a level imposed by the vertical gravitational field, $\Sigma_{\rm
  SFR} \propto W$.

In mid- and outer-disk regions (generally where \sig\ \aplt\ 100
\msunpc), the warm ($T\sim 10^4$ K) ISM is space-filling and GMCs
appear to be self-gravitating structures that do not participate in
the general vertical equilibrium.  For this regime, OML10 show that
the thermal/dynamical equilibrium theory is in good agreement with
observations, with \sigsfr\ depending on both \sig\ and the stellar
density $\rho_*$ of the disk (see also KKO11).  For outer disks,
diffuse\footnote{We use the term ``diffuse'' to refer to spatially
  dispersed gas (both warm intercloud medium and cold cloudlets) that
  does not occur in gravitationally bound molecular clouds; see
  Section 2.2 of OML10.}  HI dominates and $\Sigma_{\rm SFR} \propto
\Sigma \sqrt{\rho_*}$ because the weight of the diffuse ISM is $W
\propto \Sigma \sqrt{\rho_*}$ in this regime.  For mid-disks, gas is
concentrated in gravitationally-bound clouds (observed as GMCs) which
have relatively uniform column density, star formation efficiency, and
other properties (probably as a result of internal feedback), such
that $\Sigma_{\rm SFR} \propto \Sigma $.

In very dense regions where \sig \apgt 100 \msunpc, such as the
Galactic center and ULIRGs, UV heating is not expected to play a
strong role, and molecular gas is pervasive rather than concentrated
in effectively isolated GMCs. The transition to the ``diffuse
molecular'' starburst regime occurs where the pressure of the ISM as a
whole exceeds the pressure of isolated, bound GMCs as found in the
outer disk.  For bound or virialized GMCs with surface density
$\Sigma_{\rm GMC}\equiv M/(\pi R^2)$ that have a
gravitational-to-kinetic energy ratio of 1 to 2, the internal pressure
is $(0.5-1) G \Sigma_{\rm GMC}^2$.  The ISM as a whole must have
midplane pressure $(\pi/2) G \Sigma^2$ if equilibrium holds and gas
dominates the gravity (see Equation [\ref{wteqn}] below); from Paper I,
this pressure is primarily turbulent, driven by star formation
feedback.  The transition to the regime where molecular clouds lose
their identity (and may be destroyed by externally-driven turbulence
rather than internal feedback) therefore occurs when $\Sigma \gtrsim
(0.5-0.8) \Sigma_{\rm GMC}\sim 50-100$ \msunpc.

In the starburst regime, the theory of Paper I suggests that SN play a
key role in controlling the overall star formation rates because they
dominate the momentum injection rate to the ISM.\footnote{While
  thermal gas pressure from \ion{H}{2} regions and radiation pressure
  are likely most important in destroying individual outer-galaxy GMCs
  containing embedded clusters (because of the time delay before
  supernovae), simple estimates suggest that for the ISM as a whole,
  the momentum input/stellar mass formed is dominated by supernovae
  (see Paper I).}  Paper I presented the analytical theory, compared
the star formation rates to observations compiled in
\citet{Genzel+10}, and provided initial results from numerical models
of SN-driven turbulent feedback in a cold ISM.  According to the
theory of Paper I, $\Sigma_{\rm SFR} \propto \Sigma^2$ is expected for
most starbursts (see Equation \ref{sigsfrpred} below).  Here, we
extend Paper I to test the predictions from self-regulation over a
wide range of galaxy and feedback parameters, using time-dependent
numerical simulations.

\subsection{Simulations of Self-Regulation Due to Feedback in 
Starbursting Environments}

In this work, we model the evolution of a molecular dominated ISM using
radial-vertical simulations, including a treatment for gas motion in
the azimuthal direction.  Using a large suite of hydrodynamic models,
we focus on the role of SN driven feedback in the starburst regime,
including its relationship to other disk characteristics such as the
overall star formation rate, disk thickness, and midplane density.  A
key feature of these simulations is that the vertical dimension is
well resolved, which is important for accurately capturing the effect
of turbulence on disk thickness, as pointed out by
\citet{Shetty&Ostriker08}.  We test the sensitivity of the results to
the assumed input parameters, such as the efficiency of star formation
in dense gas, and the momentum injected per unit stellar mass.  Our
analysis aims to understand the role of feedback-induced turbulence on
the self-regulation of star formation in high (surface) density
regions, where \sig \apgt 100 \msunpc, representative of the ISM in
(U)LIRGs and galactic centers.

This paper is organized as follows.  The next section describes the
relevant equations and our numerical methods.  Section 3 presents our
model results, as well as a comparison between the simulations and the
predictions from self-regulation theory.  After a discussion we
summarize our work in Section 4.

\section{Numerical Methods} \label{methosec}

\subsection{Basic Equations and Local Disk Model}

To model the evolution of the ISM in dense molecular disks, we solve
the time-dependent hydrodynamic equations, including self-gravity.
The relevant equations are:
\begin{eqnarray}
\frac{\partial \rho}{\partial t} + \nabla \cdot (\rho {\bf v}) &=& 0 \label{heqn1} \\
\frac{\partial {\bf v}}{\partial t} + {\bf v} \cdot \nabla {\bf v} &=& -\frac{1}{\rho}\nabla P - 2{\bf \Omega}\times {\bf v} - \nabla\Phi_{g}+{\bf g_{ext}}   \\
\label{heqn2}
\nabla^2 \Phi_g&=&4\pi G \rho
\label{heqn3}
\end{eqnarray}
where $\rho$, ${\bf v}$, $P$, and ${\bf \Omega}$ are the volume
density, velocity, pressure, and angular velocity of the rotating
frame, respectively, and
$G$ is the gravitational constant.  Since gas cools efficiently in
high density molecular regions, we employ an isothermal equation of
state, with constant sound speed $c_s = (P/\rho)^{1/2}$.  We implement
a static stellar gravitational field ${\bf g_{ext}}=-\Omega^2 z
\hat{z}$ (assumed to arise from a spherical bulge), which helps to
concentrate gas to the midplane. The bulge potential is also responsible for the
overall rotation of the gas with angular velocity $\Omega$.  
The time-varying self-gravitational
potential due to the gas is $\Phi_g$.

The domain of our simulations is two-dimensional (2D), consisting of a
radial and vertical (${\bf \hat{R}}, \, {\bf \hat{z}}$) cross-section
of a galactic disk, with extents \Lr\ and \Lz, respectively.  Though
the 2D simulations only treat $x=R-R_0$ and $z$ as independent
variables, velocities in all three directions (including $\hat \phi$)
are included.  We also include Coriolis forces, with ${\bf \Omega} =
\Omega \hat z$ constant (i.e. solid body rotation, for a
constant-density bulge).  
We do not consider shear, as our focus is on
galactic central regions, where the rotation curve is still rising.
When $d\Omega/dR=0$, the tidal potential 
term in the rotating frame is zero and does not enter the momentum
equation (this tidal term is nonzero in outer-disk
regions where rotation is strongly sheared -- see the right-hand side
of Equation 15 of KKO11).

Additionally, in our calculation of star formation rates, we
implicitly consider the extent in the azimuthal direction \Lphi\ (see
Section \ref{modsec}).  
To model a local patch of the disk cross-section, we adopt periodic
boundary conditions in ${\bf \hat R}$.  In order to maintain a
constant value of \sig\ throughout the simulation, we also adopt
periodic boundary conditions in ${\bf \hat z}$.  As we describe in
Section \ref{testsec}, we ensure that \Lz\ is sufficiently large in
order to follow the complete evolution of the supernova shells, such
that the ISM scale height and star formation rate are converged.
Simulating 2D (${\bf \hat{R}}, \, {\bf \hat{z}}$) slices allows us to
perform calculations with very high (sub-pc) spatial resolution, as
well as to explore a wide range in parameter space (which may be used
as a basis for future three dimensional [3D] simulations; initial
tests show that similar results hold for 3D models).

We numerically integrate the hydrodynamic Equations (\ref{heqn1}) -
(\ref{heqn3}) using the {\it Athena} code \citep{Stone+08}.  {\it
  Athena} solves the partial differential equations using a
single-step, directionally unsplit Godunov method in multiple
dimensions \citep{Stone&Gardiner09}.  We adopt piecewise-linear
reconstruction and the HLLC Riemann solver.  To solve the time-varying
self-gravitational potential $\Phi_g$, we employ a Fourier transform
method with vacuum vertical boundary conditions and periodic
horizontal boundary conditions, as described in
\citet{Koyama&Ostriker09b}.  We explore a range in \Lr\ and \Lz, as
well as the number of zones \Nr\ and \Nz, in order to ensure that the
results are not sensitive to the domain extent and that the features
are well resolved numerically, as we discuss in Section \ref{testsec}.

\subsection{Feedback Prescription and Model Parameters}

Equations (\ref{heqn1}) - (\ref{heqn3}) only describe the basic
hydrodynamics, rotation, gas self-gravity, and the vertical potential.
Our simulations also include an idealized model of momentum feedback
produced by supernovae, which drives turbulence and disperses dense
regions.  This feedback mechanism increases the total pressure, and
limits collapse of the gaseous disk to only a small fraction of the
densest material.

Our method to identify regions that could form stars, and to apply
momentum feedback that these stars would produce, is similar to that
described in \citet{Shetty&Ostriker08}.  Here, we provide an overview
of this algorithm, and refer the reader to \citet{Shetty&Ostriker08}
and KKO11 for a more detailed description.

We employ a statistical approach to determine host locations for the
feedback events, and how much star formation is tallied (we do not
remove gas from the domain).  Star formation can occur in a fraction
of the regions where the number densities are greater than some chosen
threshold density \nth.  Thus, at every time-step each grid zone with
$n \ge$ \nth\ is identified.  Next, the number of massive stars (that
can produce feedback) in zones with $n \ge$ \nth\ is determined
through a probability defined by two other user-chosen parameters, the
``free-fall efficiency'' (conversion of gas mass to stars per
free-fall time), \epsnth, and the total mass in all stars formed per
high mass star, \msn.  We then apply feedback instantaneously, centered
on those zones where high mass stars are determined to form (i.e. we
omit time delays and spatial offsets in feedback, which more realistic
models would take into account).  The probability of a feedback event
centered on a zone with $n \ge$ \nth\ in a given time-step $\Delta t$
is thus
\begin{equation}
P = \frac{\Delta t \, \epsilon_{\rm ff}(n_{\rm th}) \,
  M_{\rm cl}}{t_{\rm ff}\,(n_{\rm th}) \, m_*},
\label{snprob}
\end{equation}
where $M_{\rm cl}$ is the mass of gas contained in the dense cloud in
which the event originates, and the free fall time is \tffnth =
$[3\pi/(32 G \mu m_p n_{\rm th})]^{1/2}$; here $\mu$ is the mean
molecular weight and $m_p$ is the proton mass.  For each massive star
formed in a given time step, the total mass in stars formed is
augmented by \msn\ (Equation 21 of KKO11).

After a zone is determined to host a supernova, a circular region with
chosen radius \rsh\ is delineated.  The density inside this region is
reset to a uniform value (conserving total mass), and velocities
pointing away from the center are set such that the mean (spherical)
momentum injected per event is equal to \psn\ (see Equation 23 of
KKO11).

\begin{deluxetable*}{cl}
   \tablecaption{Symbols Employed}

\tablehead{ 
\colhead{Symbol} & \colhead{Definition} 
}

\startdata 

\multicolumn{1}{c}{Simulation Parameters} \\ \\

\epsnth & free-fall efficiency at the threshold density \\

\Lr & physical extent in radial dimension \\
\Lz & physical extent in vertical dimension \\

\msn & total mass of stars per feedback event \\

\Nr & number of zones in radial dimension \\
\Nz & number of zones in vertical dimension \\

\nth & threshold number density for feedback to occur \\

$\Omega$ & angular velocity \\

\psn & injected momentum per feedback event \\

\rsh & radius of feedback event \\

$\Sigma$ & gas surface density \\
\torb & orbital time \\ \\

\hline \\
\multicolumn{1}{c}{Measured Quantities} \\ \\

$\epsilon_{\rm ff}(n_{\rm 0})$ & free-fall efficiency at midplane density\\
$f_p$ & turbulent dissipation parameter \\
\Ht & gas disk thickness \\
$n_0$ & gas number density at midplane \\
\pdriv & vertical momentum injection rate per unit area \\
\pturb & midplane turbulent pressure \\
\sigsfr & star formation rate \\
\vz  & vertical velocity dispersion \\
$v_z$  & vertical velocity \\
$W$ & vertical weight of the gas layer \\
$\chi$ & contribution of bulge to vertical gravity, relative to gas self-gravity \\

\enddata
{\singlespace }
\label{nottab}
\end{deluxetable*}

In summary, there are five user-defined parameters required to
identify and implement feedback: \nth, \epsnth, \rsh, \psn, and \msn.
We adopt \msn = 100 \msun, which is derived from a \citet{Kroupa01}
IMF assuming supernovae result from stars with mass $\ge$ 8 \msun.
The chosen value of \rsh\ also sets the effective azimuthal thickness
\Lphi = 2\rsh, which is used in setting $M_{\rm cl}$.  The remaining
three parameters, along with \sig, \omg, $c_s$, \Lr, \Lz, and the
resolution \Nr$\times$\Nz\ complete the set of inputs for each
numerical simulation.  Table \ref{nottab} lists the symbols and the
corresponding description of the relevant model parameters and
measured quantities we refer to throughout this paper.
  
Our initial vertical density profile decreases as a Gaussian away from
the midplane, such that the surface density is \sig.  We also include
a sinusoidal perturbation along $R$, to seed gravitational
instability.  We have verified that our particular choice of initial
conditions does not affect the later evolution in any way.  As we
demonstrate in the next section, by approximately one orbital time
\torb = 2$\pi/\Omega$, the dynamic disk settles into a statistical
steady state, such that the downward motions due to the vertical
potential are countered by the upward motions due to feedback
occurring near the midplane.

\subsection{Missing Physics}
The hydrodynamic models we consider are highly idealized, while in the real
ISM a number of additional physical processes may play a role.
Cosmic rays, magnetic fields, and thermal radiation
can contribute pressure, and can in principle affect self-regulation of 
star formation.  The first two are, however, likely to be less important 
than the turbulent pressure if cosmic ray and magnetic scale heights 
are large compared to that of the neutral disk, and the last is likely 
important only if the gas surface density is extremely high (see Paper I). 
The analytical model for self-regulated star formation in Paper I allows for 
feedback processes in addition to the turbulent driving considered here,
and it will be interesting to explore these effects quantitatively 
in future simulations.

As our simulations represent radial-vertical slices rather than full
three-dimensional regions, we cannot study the detailed morphological
structure of the ISM, such as filaments and the shapes of dense
clouds.  Three-dimensional simulations would be necessary to
characterize the masses of clouds, and to make comparisons to
structures as identified in position-position-velocity molecular-line
data cubes \citep[e.g.][]{Pichardo+00,Ostriker+01, Gammieetal03,
  Shetty+10}.  Because the interior of vertically-expanding shells can
be ``filled'' by gas moving horizontally from other azimuthal
locations, the morphology in our present simulations appears more
``open'' than it would in a fully three dimensional model.  We note,
however, that three-dimensional simulations of self-regulated star
formation in outer disks analogous to the radial-vertical 
models of KKO11 give star
formation rates that are quite consistent with those obtained using
radial-vertical simulations.

Because the primary focus of this work is on star formation in the
molecule-dominated turbulent ISM, we have adopted the same (highly
idealized) assumption of an isothermal medium that has been so
fruitful in many of the first numerical studies of turbulent molecular
clouds (see reviews by \citealt{MacLow&Klessen04} and
\citealt{McKee&Ostriker07}).  In reality, the ISM has much more
complex thermal and chemical structure, and a number of recent
numerical studies have taken these into consideration.  In particular,
three-dimensional simulations including detailed heating and cooling
for ISM models with thermal supernova energy injection have recently
been conducted by \citet{deAvillez&Breitschwerdt04, Joung+09,
  Hill+12}, among others.  Although most simulations including a hot
ISM have focused on conditions similar to the Solar neighborhood,
\citet{Joung+09} included a case with very high supernova rate, as
would be expected for star formation rate $\sim 1$ \sfrunits, similar
to the starburst regime we consider here.  These recent multiphase
simulations have not included self-gravity, however, and thus the
supernovae rate is imposed as an input parameter rather than being
modulated by the mass of gravitationally-collapsing gas.  It will be
quite interesting to include self-gravity and a feedback
implementation together with multiphase heating and cooling to model
self-regulated star formation more realistically.  In particular, by
comparison with simulations that model supernovae by injecting thermal
energy, it will be possible to assess the simple momentum injection
model we adopt here to represent turbulent driving in the neutral ISM
by radiative supernova remnants.

\section{Results \label{resultsec}}

\subsection{Overview of Simulations}\label{overviewsec}

\begin{deluxetable*}{cccccccccccc}
  \tablewidth{0pt} \tablecaption{Input Parameters of Hydrodynamic
    Models\tablenotemark{a}}

\tablehead{ 
\colhead{(1)} & \colhead{(2)} &
\colhead{(3)} & \colhead{(4)} &
\colhead{(5)} & \colhead{(6)} & \colhead{(7)} &
\colhead{(8)} \\
\colhead{Model} & \colhead{$\Sigma$} &
\colhead{\epsnth} & \colhead{\psn} &
\colhead{$\Omega$} & \colhead{\torb} & \colhead{$L_R$} &
\colhead{$L_z$} \\
\colhead{ } & \colhead{(\msunpc)} & \colhead{} &
\colhead{(\msun\ \kms)} & \colhead{(Myr$^{-1}$)} & \colhead{(Myr)} & \colhead{ (pc)} &
\colhead{ (pc)} 
}

\startdata

Series S  & (variation in $\Sigma$) \\

S100 & 100 & 0.005 & $3\times 10^5$ & 0.1 & 62.8 & 120 & 240  \\
S200 & 200 & 0.005 & 3$\times 10^5$ & 0.2 & 31.4 & 60 & 120 \\
S400 & 400 & 0.005 & 3$\times 10^5$ & 0.4 & 15.7 & 30 & 60 \\
S800 & 800 & 0.005 & 3$\times 10^5$ & 0.8 & 7.9  & 30 & 60 \\
\\

Series E &  (variation in \epsnth) \\
E0.005 & 200 &  0.005 & $3 \times 10^5$ & 0.2 & 31.4 & 60 & 120 \\
E0.01 & 200 &  0.01 & $3 \times 10^5$ & 0.2 & 31.4 & 60 & 120 \\
E0.025 & 200 &  0.025 & $3 \times 10^5$ & 0.2 & 31.4 & 60 & 120 \\
E0.05 & 200 &  0.05 & $3 \times 10^5$ & 0.2 & 31.4 & 120 & 240 \\
\\

Series PA &  (variation in \psn) \\
PA1.5 & 100 &  0.005 & $1.5 \times 10^5$ & 0.1 & 62.8 & 120 & 240 \\
PA3 & 100 &  0.005 & $3 \times 10^5$ & 0.1 & 62.8 & 120 & 240 \\
PA6 & 100 &  0.005 & $6 \times 10^5$ & 0.1 & 62.8 & 120 & 240 \\
PA9 & 100 &  0.005 & $9 \times 10^5$ & 0.1 & 62.8 & 120 & 240 \\
\\

Series PB & (variation in \psn) \\
PB1.5 & 200 &  0.01 & $1.5 \times 10^5$ & 0.2 & 31.4 & 60 & 120 \\
PB3 & 200 &  0.01 & $3 \times 10^5$ & 0.2 & 31.4 & 60 & 120 \\
PB6 & 200 &  0.01 & $6 \times 10^5$ & 0.2 & 31.4 & 120 & 240 \\ 
PB9 & 200 &  0.01 & $9 \times 10^5$ & 0.2 & 31.4 & 120 & 240 \\ 
\\

Series O &  (variation in $\Omega$) \\
O1 & 200 &  0.005 & $3 \times 10^5$ & 0.1 & 62.8 & 60 & 120 \\ 
O2 & 200 &  0.005 & $3 \times 10^5$ & 0.2 & 31.4 & 60 & 120 \\ 
O4 & 200 &  0.005 & $3 \times 10^5$ & 0.4 & 15.7 & 60 & 120 \\ 
O8 & 200 &  0.005 & $3 \times 10^5$ & 0.8 & 7.9 & 60 & 120 \\ 

\enddata
{\singlespace 
\tablenotetext{a}{All listed models have $N_R \times$ $N_z$ = 512 $\times$ 1024 zones.}
}
\label{simtab}
\end{deluxetable*}

We have explored a large range in simulation parameters in order to
develop a robust understanding of the effects of momentum feedback in
high density, rotating disks.  Table \ref{simtab} lists the main
simulations we consider here.  We classify the simulations into five
groups, based on the parameters which are varied.  Column (1)
indicates the name of each simulation, as well as the group to which
it belongs.  Columns (2) - (6) list the input values of surface
density \sig, star formation efficiency per free-fall time at the
threshold density \epsnth, momentum injected per supernova \psn,
rotational speed $\Omega$, and orbital time \torb, respectively.  The
last two columns list the $R$ and $z$ dimensions of the simulation
domain.  Notice that some models are repeated in different Series:
S100 = PA3, S200 = E0.005 = O2, and E0.01 = PB3.  We further note that
although we have executed and analyzed well over 100 additional
simulations, those listed in Table \ref{simtab} span a sufficiently
broad range of the parameters to highlight the major findings of our
research.

We have also explored variations in the other parameters required to
execute the simulations: \msn, $c_s$, \nth, \rsh, \Lr, \Lz.  As we
discuss, \msn\ always occurs as a ratio with \psn\ in the relevant
equations, so any variation in \psn\ is equivalent to a corresponding
variation in \psn/\msn.  We thus fix \msn\ to 100 \msun, and vary
\psn.  We vary $p_*$ about the value expected for a supernova that has
reached the shell formation stage \citep[e.g.][]{Blondin+98}:
\begin{equation}
p_* \sim 3\times 10^5 M_\odot \, {\rm km\, s^{-1} }
\left(\frac{E_{\rm SN}}{10^{51}{\rm erg}}\right)^{0.94}
\left(\frac{n_0}{{\rm cm}^{-3}} \right)^{-0.12};
\label{SNfiducial}
\end{equation}
this is insensitive to the ambient density $n_0$ and approximately
linear in the supernova energy.
In all the simulations, we set $c_s$=2 \kms.  This value is larger
than the sound speed of cold (T\aplt 100 K) gas.  Such values are
necessary because without magnetic fields, shocked gas would result in
unrealistically high density regions, and thus lead to very small
time-steps in the numerical simulations.  Since turbulent motions
still dominate, and to partly account for these (unmodeled) magnetic
effects, we set $c_s$=2 \kms.  We have verified that provided $c_s$ is
small compared to the turbulent velocity, the precise value does not
significantly affect the results.  We also note that for the analogous
simulations of KKO11, initial tests show that inclusion of magnetic
fields do not significantly alter the results.  For the remaining
input parameters \nth, \rsh, and box size \Lr, \Lz, we discuss 
effects on the disk evolution in the subsequent sections.

\subsection{Box Size and Resolution Tests}\label{testsec}

Before presenting the simulation results, we verify that the choice of
domain size and the numerical resolution do not affect the outcome.
Since we employ periodic boundary conditions, the extent in $z$ must
be large enough such that gas flow across the $z$ boundary is
unimportant.  Gas leaving the (top or bottom) $z$ boundary returns
through the opposite boundary; if outflow velocities were large, there
would be a corresponding spurious compression of gas toward the
midplane by the returning inflow.  By constructing sufficiently large
vertical domains, we ensure that there is little mass and momentum
flux through the boundaries.\footnote{In reality, hot gas produced by
  supernovae and high-altitude material accelerated by radiation
  forces may escape as a wind; the current simulations focus on cold,
  dense gas and do not include these effects.}  In addition to the
size of the domain, the physical resolution must be sufficiently high
to ensure that any gaseous structures that form, such as the high
density clouds, are well resolved so that the Truelove criterion is
satisfied \citep{Trueloveetal97}.

\begin{figure*}
\includegraphics[width=18cm]{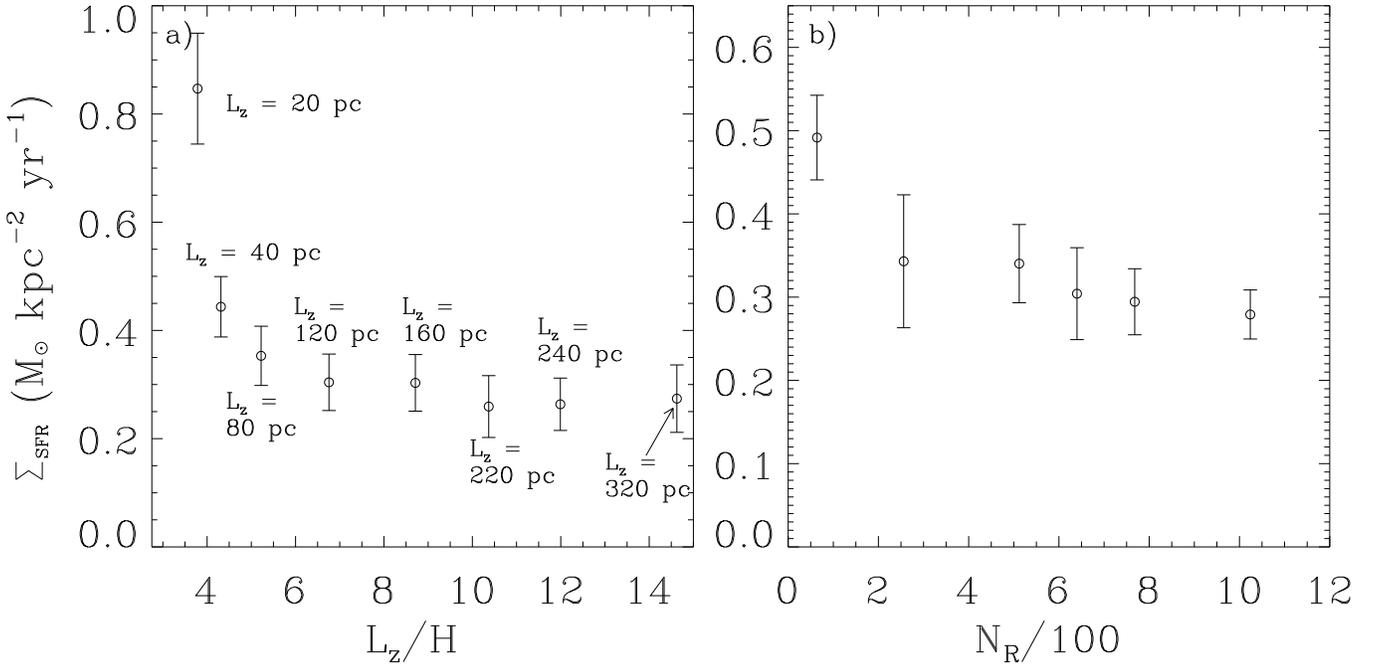}
\caption{a) The effect of box size \Lz$/H$ on \sigsfr.  Points show
  the mean \sigsfr\ from model PB6, but with different extents in \Lz.
  b) \sigsfr\ from S200 models with different resolutions (\Nz\ = $2
  \times$\Nr).  In both panels, the points correspond to mean values
  of \sigsfr\ in 10 Myr bins, beyond 50 Myr from the start of the
  simulation.}
\label{restest}
\end{figure*}

Figure (\ref{restest}a) shows how the steady-state \sigsfr\ (defined
in next subsection) depends on box size for model PB6, for a given
physical resolution.  When the ratio of \Lz\ to the disk thickness
\Ht\ (also defined below) is small, the midplane density is
artificially enhanced (as described above), triggering more cloud
collapse and subsequent supernova explosions.  The star formation rate
decreases as \Lz/$H$ increases, with fewer shells passing through the
boundary. At large \Lz/$H$, \sigsfr\ converges to a limiting value.

The momentum fluxes ($\rho v_z^2$) through the top and bottom
boundaries are $\sim$43\% of the momentum flux in the disk midplane
for the simulation with \Lz=20 pc.  In the model with \Lz=320 pc, the
boundary momentum flux is \aplt 0.5\% of the value at the midplane.
Similarly, we measure large time-averaged vertical mass flows
$\langle\rho |v_z|\rangle$ at the vertical boundaries in models with
insufficient extents.\footnote{The time-averaged true mass flux $\langle\rho
  v_z\rangle$ is zero at all heights.}  For the model with \Lz=20 pc,
the ratio of $\langle\rho |v_z|\rangle$ at the vertical boundaries to
that in the midplane is \apgt 0.6.  When the vertical extent is
sufficiently large, such as the model with \Lz=320 pc, this ratio is
\aplt0.01.  Based on a large number of tests of different models, we
have found that the ratio of vertical mass flow and momentum flux in
the vertical boundary to the corresponding value in the disk midplane
is negligible when \Lz/$H$ is large.  Correspondingly, we find that
\sigsfr\ converges provided \Lz/$H $ \apgt 6, so that for all
simulations we choose a domain size such that \Lz/$H>$ 6, for the
measured value of $H$.

To ensure that the simulation results are independent of numerical
resolution, we have executed a number of simulations with the same box
size, initial conditions, and feedback parameters, but with varying
\Nr\ and \Nz.  Figure (\ref{restest}b) shows the mean value of
\sigsfr\ for the fiducial model S200 with different resolutions, all
with box size \Lr \tm \Lz = 60\tm 120 \pcsq.  Clearly,
\sigsfr\ converges to within 15\% for all cases with dimension \Nr \tm
\Nz\ $>$ 256\tm 512.  For \Nr\tm\Nz\ = 256\tm 512, the physical
resolution in model S200 is 0.23 pc; at our standard size \Nr\tm\Nz\ =
512\tm 1024, the physical resolution is 0.12 pc.  Our largest box is
twice as large as that of model S200, with resolution 0.23 pc.  At
this resolution, the highest density at which the Truelove criterion
($\lambda_J/4 > L_z/N_z $, for Jeans length $\lambda_J =
c_s[\pi/(G\rho)]^{1/2}$) is satisfied is $\sim 10^5$ \cmt, whereas
typical cloud densities in our models are \aplt $10^4$ \cmt.  Thus, in
order to explore a large range of parameters, and at the same time be
confident that the simulations are sufficiently well resolved, we will
employ \Nr\tm\Nz\ = 512\tm 1024 as the standard resolution.

We have also explored the impact of the remaining user defined
parameters, \rsh, \nth, and \Lr.  For ambient density of $\sim
100-1000$ \cmt\ (similar to mean densities in our models), supernova
remnants become radiative when their radii are a few pc \citep[e.g.][
  Equation 39.21]{Drainebook11}.  We adopt a standard value of \rsh = 5
pc, and find similar simulation behavior for any other \rsh\ within a
factor 2 of this value.  We find that when \nth \apgt 5000 \cmt, the
evolution is not strongly dependent on the choice of \nth.  

Because we have periodic boundary conditions in the radial direction,
the value of $L_R$ does not affect the evolution provided that 
$L_R\apgt H$ and that the time-averaged gas distribution remains
uniform and ``disk-like'' in the radial
direction.  However, for some conditions the
value of the Toomre $Q$ parameter (using the turbulent velocity
dispersion) will be small enough that the combination of turbulence
and rotational support is insufficient to prevent radial collapse
under self-gravity for large radial domains (see Equation 29 of Paper I).
As discussed in Appendix B of Paper I, the massive structures that
form as a consequence of this collapse in real galaxies may
potentially be dispersed by radiation pressure.  However, in the
current work we have not implemented radiation forces, so we consider
only models that do not lead to overall radial collapse.  For our
fiducial model S200, the value of $Q$ is less than unity, so that
collapse ensues if we use a large radial domain.  However, we have
confirmed that if we increase $\Omega$ by a factor 1.5 (with other
parameters as in the S200 model) such that $Q>1$, a model with $L_{\rm
  R}=120$ pc is in all respects quite similar to the same model run
with $L_{\rm R}=60 {\rm pc}$; e.g. $\Sigma_{\rm SFR}$ differs by only
$\sim 10\%$.  Additionally, we considered a model similar to S200 but
with \omg=0.4 Myr$^{-1}$ and \psn=6$\times10^5$ \msun\ \kms\ such that
the Toomre parameter is in the stable regime.  We find that the models
with \Lr\ between 50 and 100 pc achieve convergence in \sigsfr\ to
within $\sim5\%$

\subsection{Model Evolution and Statistical Properties}\label{modsec}

\begin{figure*}
\includegraphics[width=18.5cm]{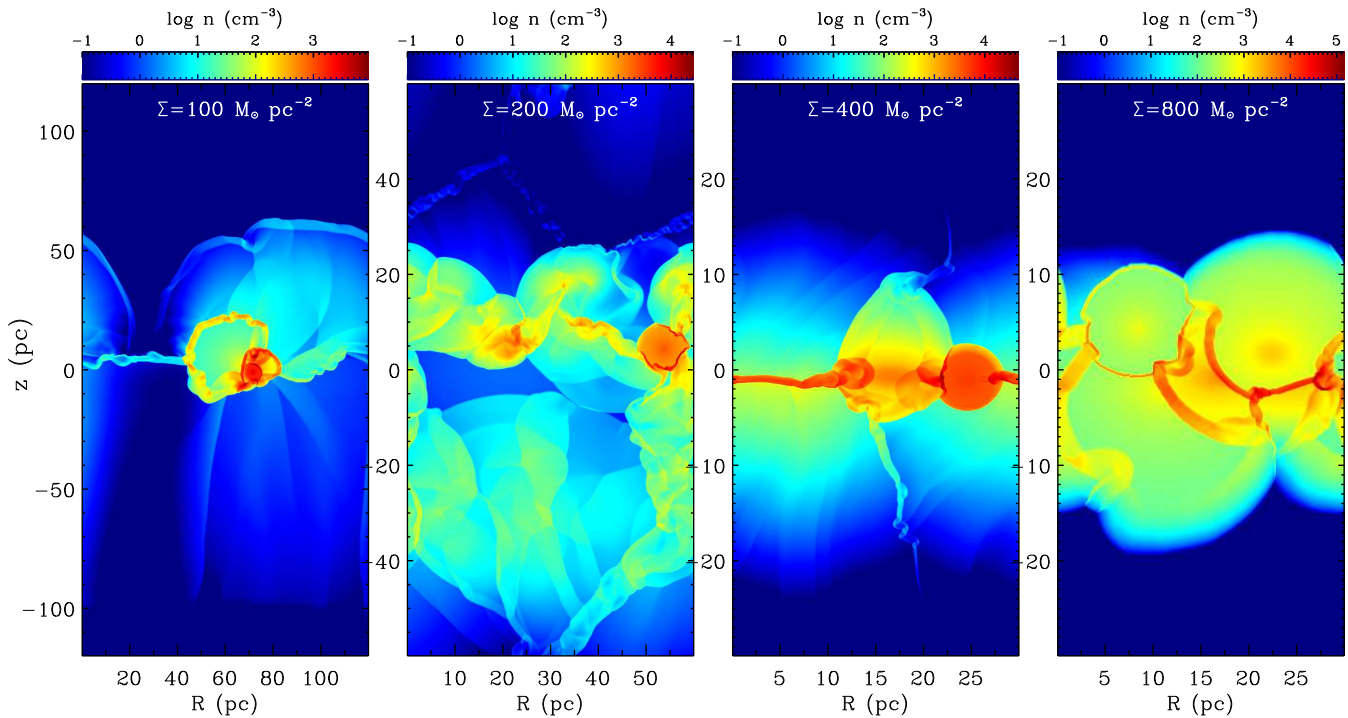}
\caption{Density of Series S models at times $t/t_{\rm orb}\approx$2.}
\label{snaps}
\end{figure*}

Figure \ref{snaps} shows the volume densities of the four Series S
models at two orbital times (2\torb) from the start of the simulation.
As we discuss below, each model approaches a statistical equilibrium
well before \torb: the star formation rate, vertical velocity
dispersion, disk thickness, and other dynamical-state parameters all
approach quasi-steady values.  Numerous evolved SN shells are evident
in Figure \ref{snaps}.  One clear trend in Series S is that in models
with higher gas surface density, the gas is also more concentrated
towards the midplane (note that the panels have different dimensions).

\begin{figure*}
\includegraphics[width=18cm]{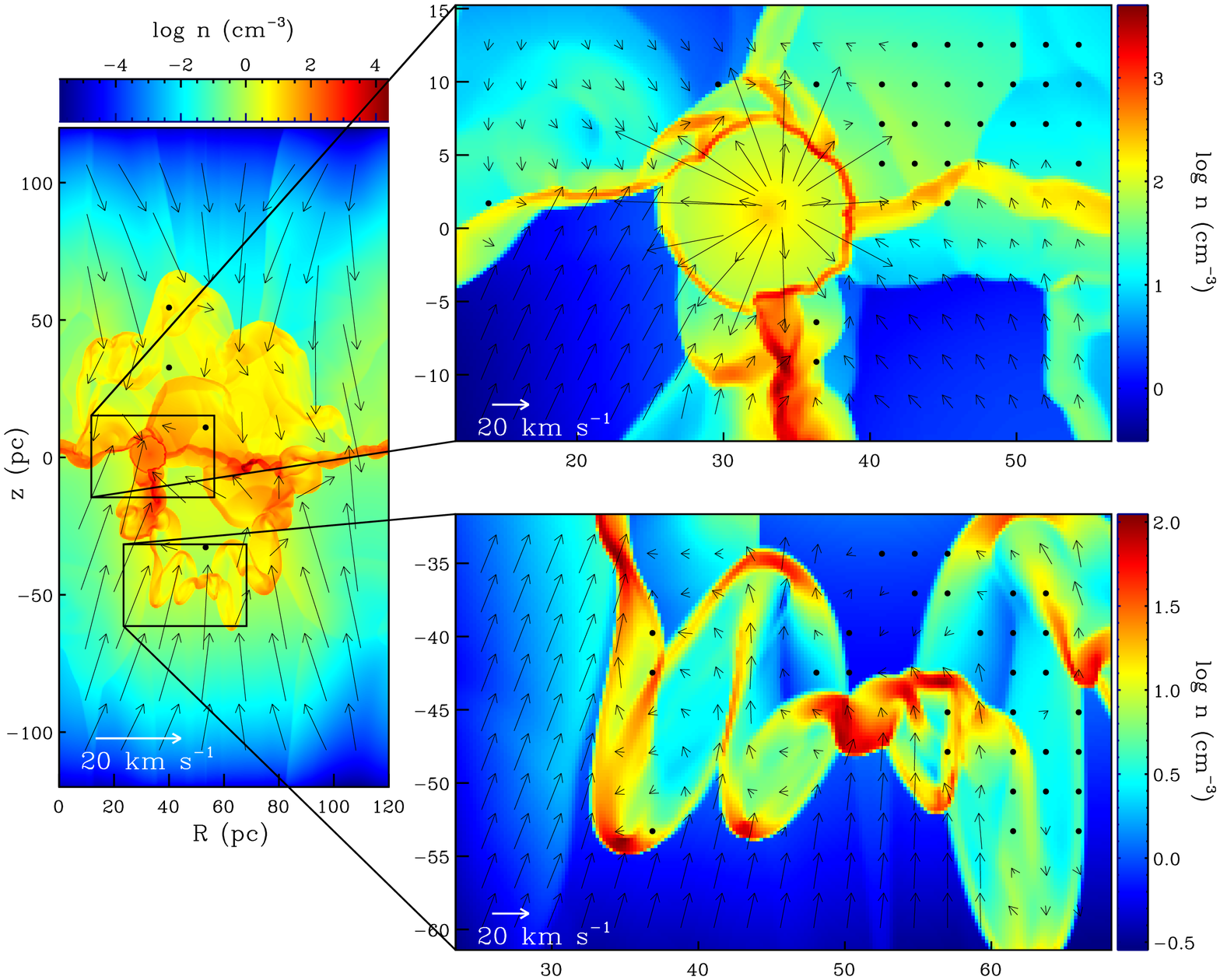}
\caption{Densities of model S100 at time $t=2.7t_{\rm orb}$=170 Myr.
  Vectors show radial-vertical velocities $(v_R^2+v_z^2)^{1/2}$.
  White vectors displayed in the bottom left of each panel show the
  vector scale.  Dots indicate locations where the velocity is $<$ 3
  \kms.  The large box is $120 \,{\rm pc}\times 240 \, {\rm pc}$, and
  the inset boxes are each $45 \,{\rm pc}\times 29 \,{\rm pc}$.  }
\label{snaps_vel}
\end{figure*}

The SN feedback events occur in the dense gas near the midplane, and
are responsible for pushing gas to higher altitudes, as well as
driving turbulence (both horizontal and vertical motions) and creating
the filamentary features easily identifiable in Figure \ref{snaps}.
Figure \ref{snaps_vel} shows the density of model S100 at $t=2.7
t_{\rm orb}$ = 170 Myr, along with the velocities in the $R, z$ plane.
The large scale velocities are generally directed towards the midplane
at this particular instant (although at other times the overall flow
is expanding, e.g. see \citealt{Walters&Cox01}).

A close-up of two regions shows the detailed density and velocity
structure.  One region focuses on a patch in the midplane where a SN
has recently exploded.  The vector field illustrates how gas within
the SN shell is rapidly expanding away from the center of the bubble,
even while surrounding gas is converging.  The other close-up shows a
region away from the midplane.  The dense regions and filamentary
structures evident here were created by interactions of gas driven by
numerous earlier feedback events.  Gas velocities near these dense
structures deviate from the large scale converging flow towards the
disk midplane.  Feedback events thus influence gas motions far from
their origin, driving turbulence throughout the simulation domain.

In each model, the star formation rate \sigsfr\ at time $t$ is
computed from the number of feedback events \Nsn\ occurring over time
interval \dtbin\ centered on $t$.  The contribution of mass to
\sigsfr\ is \Nsn\msn, where \msn\ is the mass of all stars formed per
star capable of undergoing a supernova.  Since the star formation
probability assumes an effective thickness of our simulation slice
\Lphi = 2\rsh, the same effective thickness is used in computing the
area of the domain projected on the horizontal plane, \Lr \Lphi.
Thus, \sigsfr\ over a given time interval is
\begin{equation}
\Sigma_{\rm SFR} = \frac{m_{*} N_{\rm SN}}{L_{\rm R} L_{\rm \phi} \Delta t_{\rm bin}}.
\label{csfr}
\end{equation}

\begin{figure}
\includegraphics[width=9cm]{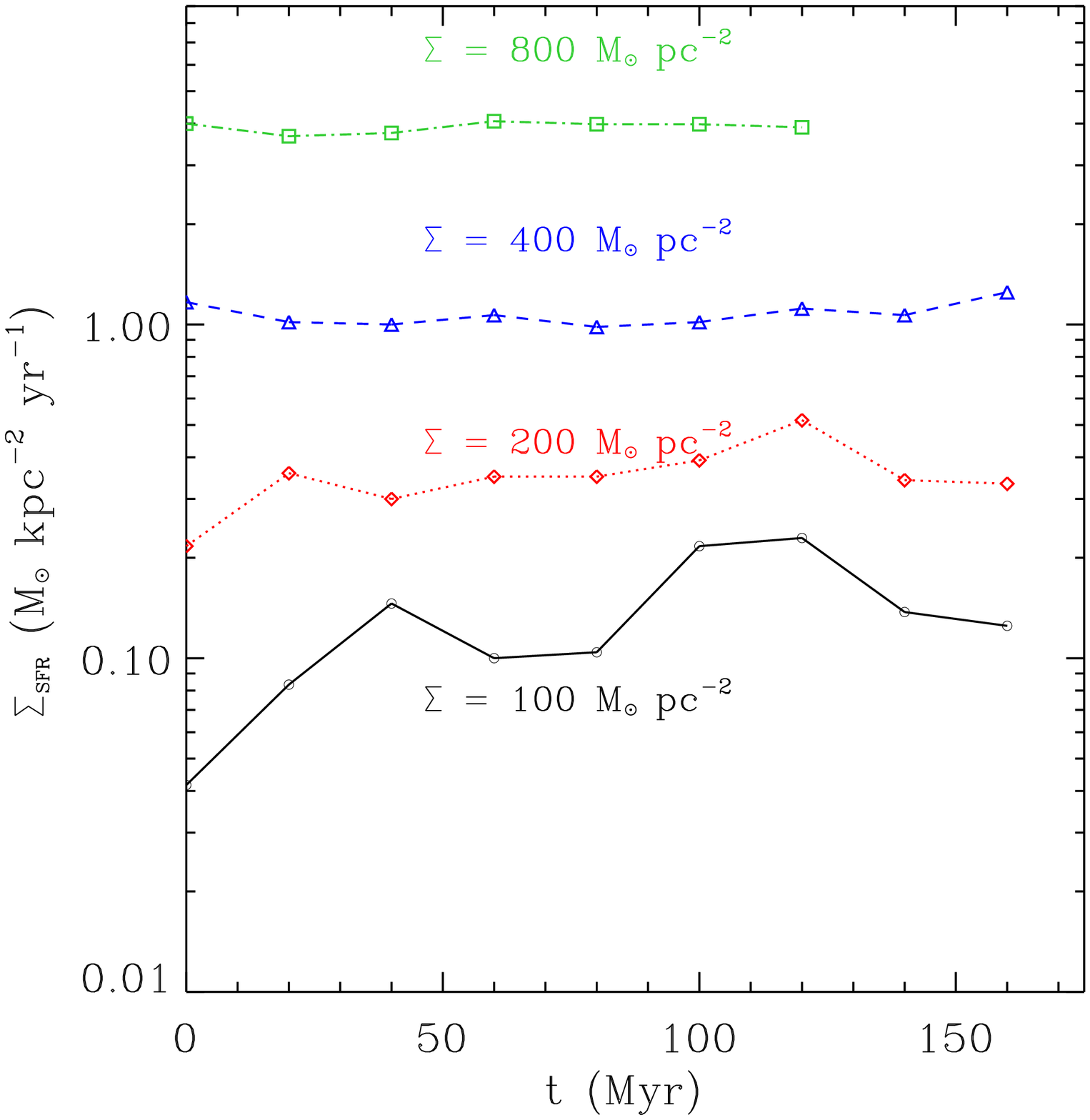}
\caption{The evolution of \sigsfr\ with time for Series S models with
  gas surface densities \sig= 100 (black circles), 200 (red diamonds),
  400 (blue triangles), and 800 (green squares) \msunpc.  Points show
  \sigsfr\ in temporal bins of \dtbin=20 Myr, as computed from
  Equation (\ref{csfr}).}
\label{sfrt}
\end{figure}

Figure \ref{sfrt} shows the evolution of \sigsfr, computed in bins of
\dtbin=20 Myr, as a function of time from all the Series S models.
This value of \dtbin\ is much larger than the vertical crossing time
of each simulation, which is simply the thickness $H$ of the disk
divided by the characteristic vertical velocity $v_z$, both of which
are defined and analyzed below.  We can thus be sure that the
estimated \sigsfr\ is averaged over a sufficiently long time such that
(on average) gas has cycled between the mid-disk $z=0$ and
out-of-plane $|z| > 0$ locations numerous times.  Figure \ref{sfrt}
indicates that \sigsfr\ saturates within 50 Myr, and as we discuss
below in Section \ref{compsec}, the saturated value generally
approaches the predictions from self-regulation.

\begin{figure*}
\includegraphics[width=15cm]{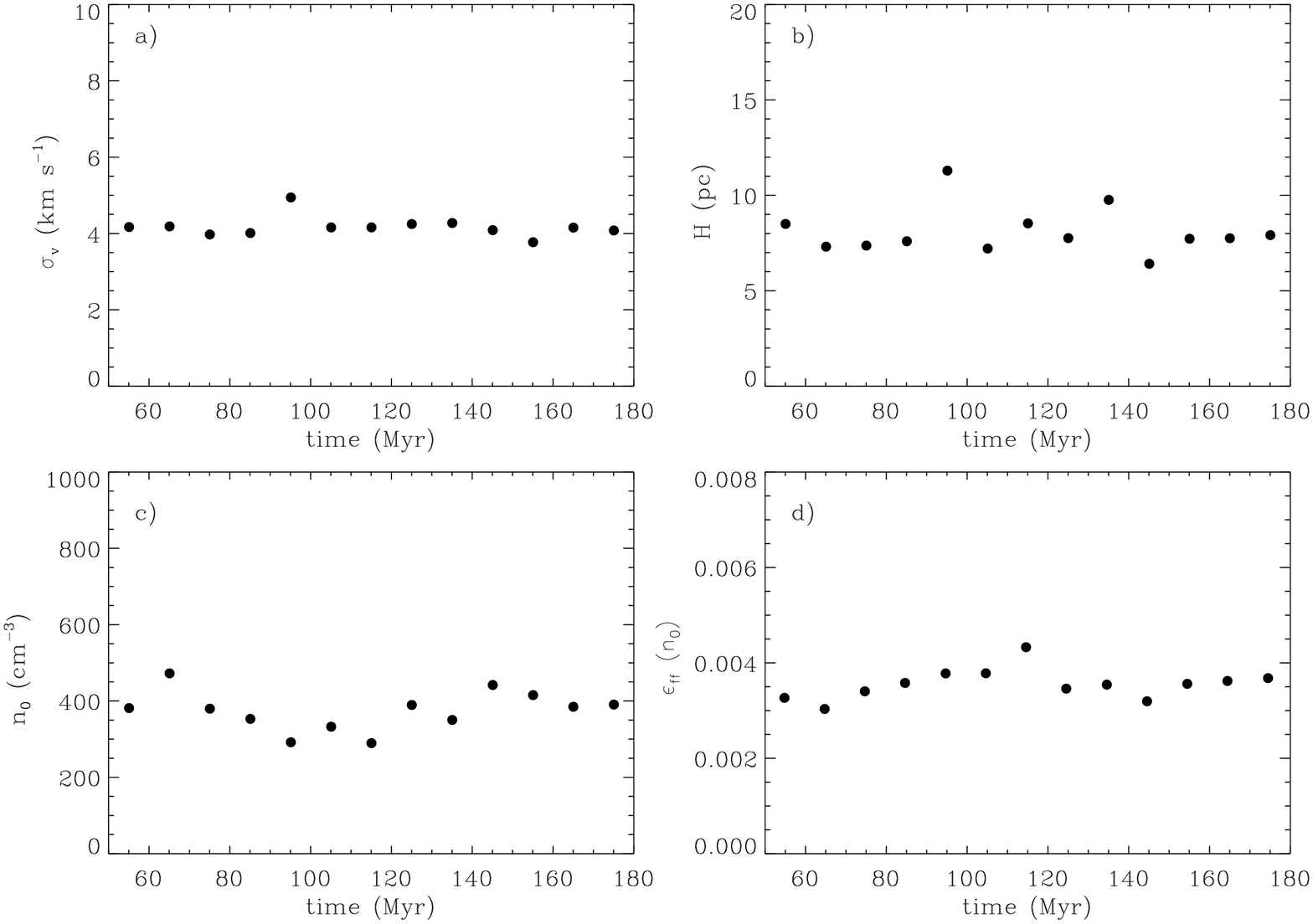}
\caption{The evolution of (a) vertical velocity dispersion \vz\ (see
  Equation \ref{wvz}), (b) gas disk thickness $H$ (see Equation \ref{wh}) (c)
  mean midplane density \no, and (d) mean star formation efficiency
  \epsrhoo\ (see Equation \ref{epscalc}), for the S200 model, averaged in
  10 Myr temporal bins.}
\label{fidpars}
\end{figure*}

Two quantities describing the gas kinematics and disk structure are
the velocity dispersion and disk thickness, respectively.  To quantify
the vertical motions, we compute the mass-weighted $\hat{z}$-velocity
dispersion through
\begin{equation}
\sigma_{v} \equiv 
\left[
\frac{\sum \rho v_z^2}{\sum \rho}
\right]^{1/2}, 
\label{wvz}
\end{equation}
where the summation is taken over all zones in the simulation.  Figure
(\ref{fidpars}a) shows the evolution of the velocity dispersion for
model S200.  As does \sigsfr, \vz\ also statistically converges, in
this case to $\sim$4.5 \kms.  Similarly, the mass-weighted disk
thickness is defined as
\begin{equation}
H \equiv \langle z^2 \rangle^{1/2} = 
\left[
\frac{\sum \rho z^2}{\sum \rho}
\right]^{1/2}.
\label{wh}
\end{equation}
Higher surface (and volume) densities lead to thinner disks, as
evident in Figure \ref{snaps}.  Figure (\ref{fidpars}b) shows that $H$
for model S200 saturates at $\sim$ 9 pc.  

Given the vertical velocity dispersion and thickness, the vertical
dynamical time is $t_{\rm ver} = H/\sigma_{z} \approx$ 2 Myr for model
S200.  The measured quantities in Figure \ref{fidpars} are the mean
values in 10 Myr bins, so that each bin corresponds to $\sim 5$
dynamical times.  Again, this allows sufficient time for gas to cycle
between the dense and diffuse phases.

Another quantity of interest is \epsrhoo, the efficiency of star
formation per free-fall time, where \tffno\ is evaluated at the
mean midplane density $n_0$.  As discussed in Paper I, \epsrhoo\ represents
the overall efficiency of star formation at the prevailing ISM
conditions, and need not be the same as the value \epsnth\ imposed to
set the rate of star formation in very high density gas (see Sections
\ref{compsec} and \ref{discsec}).

Since the star formation rate can be directly measured through
Equation (\ref{csfr}), and \tffno\ can be calculated from the
(horizontally- and time- averaged) midplane density 
\no\ measured in the simulations, the mean measured star formation efficiency
is given by:
\begin{equation}
\epsilon_{\rm ff}(n_{\rm 0}) \equiv \Sigma_{\rm SFR} t_{\rm ff}(n_{\rm 0})/\Sigma.
\label{epscalc}
\end{equation}
Figure (\ref{fidpars}c) and (\ref{fidpars}d) respectively show the
evolution of the midplane density, \no, and the mean efficiency,
\epsrhoo, for model S200.  As with \vz\ and \Ht, these quantities also
saturate, with steady-state values \no=385 \cmt\ and \epsrhoo=0.0041.

\begin{figure*}
  \includegraphics[width=15cm]{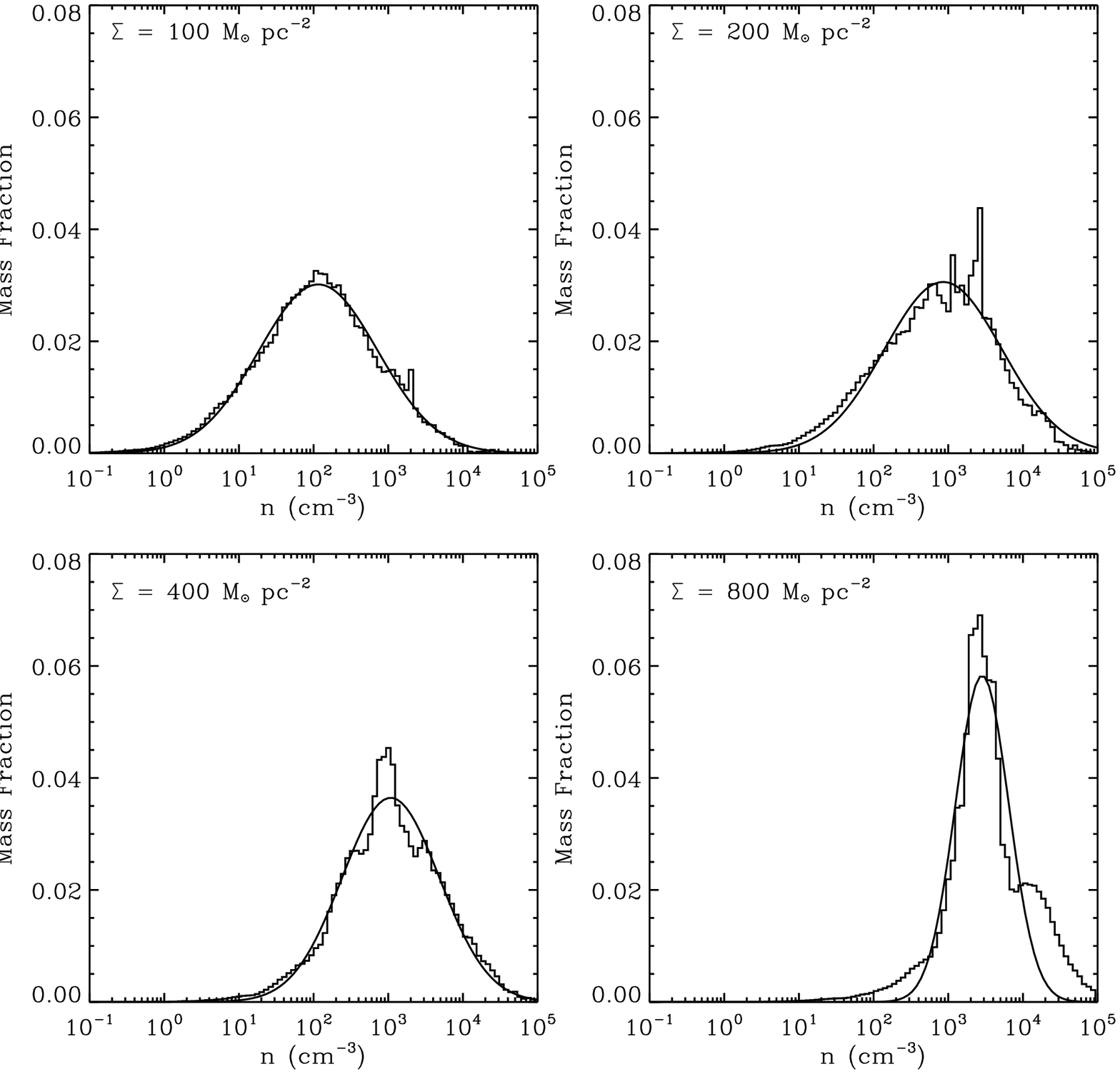}
\caption{Mass-weighted density PDFs for Series S models.  The
  histograms show the average PDFs of densities at 5 Myr intervals,
  from times 50 Myr $< t < $ 150 Myr.  The solid lines show the best
  fit log-normal distribution.  The means (standard deviations) for
  these mass-weighted log($n$) histograms in models S100, S200, S400,
  and S800 are, respectively, 2.04 ($\pm$0.81), 2.83 ($\pm$0.80), 3.02
  ($\pm$0.69), and 3.51 ($\pm$0.55).}
\label{denpdf}
\end{figure*}

\begin{figure*}
  \includegraphics[width=15cm]{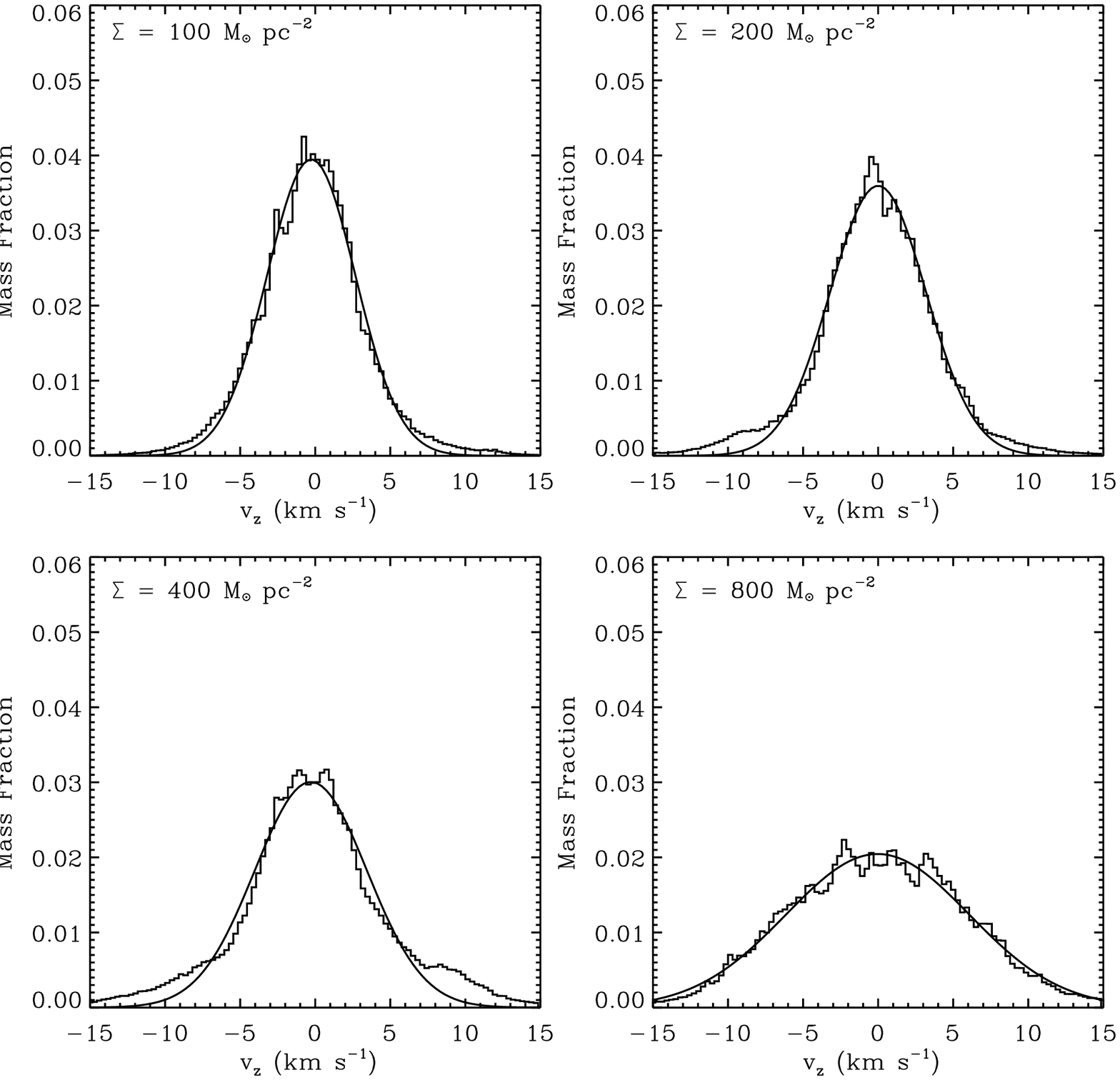}
\caption{Mass-weighted vertical-velocity PDFs for Series S models.
  The histograms show the mass-weighted velocity distributions taken
  at 5 Myr intervals, from times 50 Myr $< t < $ 150 Myr.  The solid
  lines show the best fit normal distribution.  The means (standard
  deviations) of these mass-weighted $v_z$ histograms in models S100,
  S200, S400, and S800 are, respectively, -0.24 ($\pm$3.45), -0.12
  ($\pm$3.92), -0.10 ($\pm$4.82), -0.03 ($\pm$5.67) \kms.}
\label{velpdf}
\end{figure*}

Figures \ref{denpdf} and \ref{velpdf} show the mass-weighted density
and velocity probability distribution functions (PDFs) for Series S
models.  The distributions show the average PDFs from times 50 Myr $<
t < $ 150 Myr, assessed in 5 Myr intervals.  The simulations with
higher surface densities produce PDFs which are systematically shifted
towards larger volume densities.  For model S800, the magnitude of the
(self-gravitational and external) potential strongly confines gas to
the disk midplane, such that the disk thickness becomes comparable to
our chosen value of the SN shell radius $\approx$5 pc (see
Fig. \ref{snaps}).  The feedback events produce thin shells of shocked gas
that have very
high densities, which result in the high-density secondary peak.  Yet,
most of the mid-disk has density \aplt\ 5000 \cmt, corresponding to
the main peak in Figure \ref{denpdf}d.  Apart from the S800 model,
these density PDFs are all well represented as log-normals, as
expected for highly compressible turbulent flows
\citep[e.g.][]{Vazquez-Semadeni94,Klessen00,Ostriker+01}.  The
mass-weighted velocity PDFs are approximately normal, but have more
pronounced tails at both high and low velocities.  The velocity PDFs
do not show any significant differences among the Series S models,
indicating that turbulent velocities are not strongly dependent on
\sig\ (or \sigsfr), a point we return to in Section \ref{compsec}.

\begin{deluxetable*}{cccccccc}
  \tablewidth{0pt} \tablecaption{Characteristics of Hydrodynamic Models\tablenotemark{a}}

\tablehead{ 
\colhead{Model} & \colhead{\sigsfr} & \colhead{\vz}  & \colhead{\Ht} & \colhead{$  n_0 $} & \colhead{$\epsilon_{\rm ff}(n_{\rm 0} )$} & \colhead{$f_p$} & \colhead{$\chi$}   \\
\colhead{ } & \colhead{(\sfrunits)} & \colhead{(\kms) } & \colhead{(pc)} & \colhead{(\cmt)} 
}

\startdata

Series S & (variation in $\Sigma$) \\

S100 & 0.15 & 4.0 & 11 & 161 & 0.0051 & 0.65 & 0.054 \\
S200 & 0.38 & 4.5 & 8.8 & 385 & 0.0041 & 1.1 & 0.069 \\
S400 & 1.1 & 5.2 & 6.5 & 861 & 0.0039 & 1.5 & 0.090  \\
S800 & 4.0 & 5.1 & 4.4 & 2157 & 0.0045 & 1.6 & 0.085  \\ 
\\

Series E & (variation in \epsnth) \\
E0.005 & 0.38 & 4.5 & 8.8 & 385 & 0.0041 & 1.1 & 0.069 \\
E0.01 & 0.28 & 5.3 & 11 & 278 & 0.0037 & 1.5 & 0.093 \\ 
E0.025 & 0.36 & 6.6 & 15 & 177 & 0.0058 & 1.2 & 0.13 \\ 
E0.05 & 0.58 & 7.9 & 20 & 140 & 0.012 & 0.79 & 0.19 \\ 

\\

Series PA & (variation in \psn) \\
PA1.5 & 0.34 & 2.6 & 4.5 & 232 & 0.0097 & 0.56 & 0.024 \\
PA3 & 0.15 & 4.0 & 11 & 161 & 0.0051 & 0.65 & 0.054 \\
PA6 & 0.031 & 6.4 & 26 & 60.5 & 0.0017 & 1.7 & 0.13 \\
PA9 & 0.022 & 8.0 & 31 & 51.5 & 0.0013 & 1.8 & 0.19 \\
\\

Series PB & (variation in \psn) \\
PB1.5 & 0.86 & 2.8 & 4.6 & 516 & 0.0081 & 0.90 & 0.027 \\ 
PB3 & 0.28 & 5.3 & 11 & 278 & 0.0036 & 1.5 & 0.093 \\
PB6 & 0.30 & 9.0 & 23 & 126 & 0.0055 & 0.82 & 0.22 \\
PB9 & 0.17 & 12 & 35 & 76 & 0.0040 & 1.1 & 0.37 \\

\\

Series O & (variation in $\Omega$) \\
O1 & 0.25 & 4.6 & 9.5 & 352 & 0.0028 & 1.5 & 0.019 \\
O2 & 0.38 & 4.5 & 8.8 & 385 & 0.0041 & 1.1 & 0.069 \\
O4 & 0.45 & 4.6 & 7.4 & 420 & 0.0046 & 1.1 & 0.24 \\
O8 & No collapse/feedback& \\
\enddata
{\singlespace 

\tablenotetext{a}{The values listed in this Table are depicted in
  Figures \ref{sfrpred_all}-\ref{epspredfig}, where the 1$\sigma$
  variations are also provided.}}

\label{paramtab}
\end{deluxetable*}

Table \ref{paramtab} summarizes the mean values \sigsfr, \vz\, \Ht,
\no\, and \epsrhoo\, for all models.  Averages are based on bins of 20
Myr, starting at $t=$ 50 Myr.  The last two columns give \fup\ and
$\chi$, quantities relevant for the analytical expressions derived in
Paper I and discussed below.  All models with entries listed for the
measured parameters reach a steady state.  However, Model O8 did not
collapse to reach high densities; we believe this is because it is
stabilized by a high rotation rate (see Sec. \ref{compsec}).  In the
following section, we compare the properties of each of the
simulations presented in Table \ref{paramtab} to the analytical
predictions from self-regulation derived in Paper I.

\subsection{Comparison with Predictions from Self-Regulation}\label{compsec}

As discussed in Paper I, in self-regulated starburst regions, the
vertical weight $W$ of the molecular disk is expected to be balanced
mainly by turbulent pressure \pturb\ (unless the optical depth to IR
exceeds $\sim 16$).  Under this framework, turbulence is driven
predominantly by feedback from massive stars, so the momentum
injection rate determines \pturb. The total upward momentum per unit
time per unit area is then given by $f_p P_{\rm drive}$ for $P_{\rm
  drive}$ a fiducial momentum injection rate per unit area associated
with star formation and $f_p$ an order-unity dimensionless constant.
Each of the gravitational, turbulent, and feedback momentum-injection
fluxes may be measured directly in the simulations through:
\begin{equation}
W = \frac{\pi G \Sigma^2}{2} (1 + \chi), 
\label{wteqn}
\end{equation}
\begin{equation}
P_{\rm turb} = \rho_0 \sigma_{v}^2,
\label{pturb}
\end{equation}
\begin{equation}
P_{\rm drive} = \frac{1}{4} \frac{p_*}{m_*}\Sigma_{\rm SFR},
\label{pdrive}
\end{equation}
respectively.  As discussed below, $\chi$ accounts for the gravity of
the stellar bulge relative to gas self-gravity, and is usually small.
In choosing a fiducial value for \psn\ we assume that radiative
supernova shells dominate the momentum injection (see Equation
\ref{SNfiducial}), but other terms could equally well be included in
Equation (\ref{pdrive}), and we explore a range of $p_*$.  If the disk
evolves to be turbulence-dominated and governed by star formation
self-regulation, then we should find that \pturb $\approx$ \pdriv
$\approx W$.

\begin{figure*}
  \includegraphics[width=18cm]{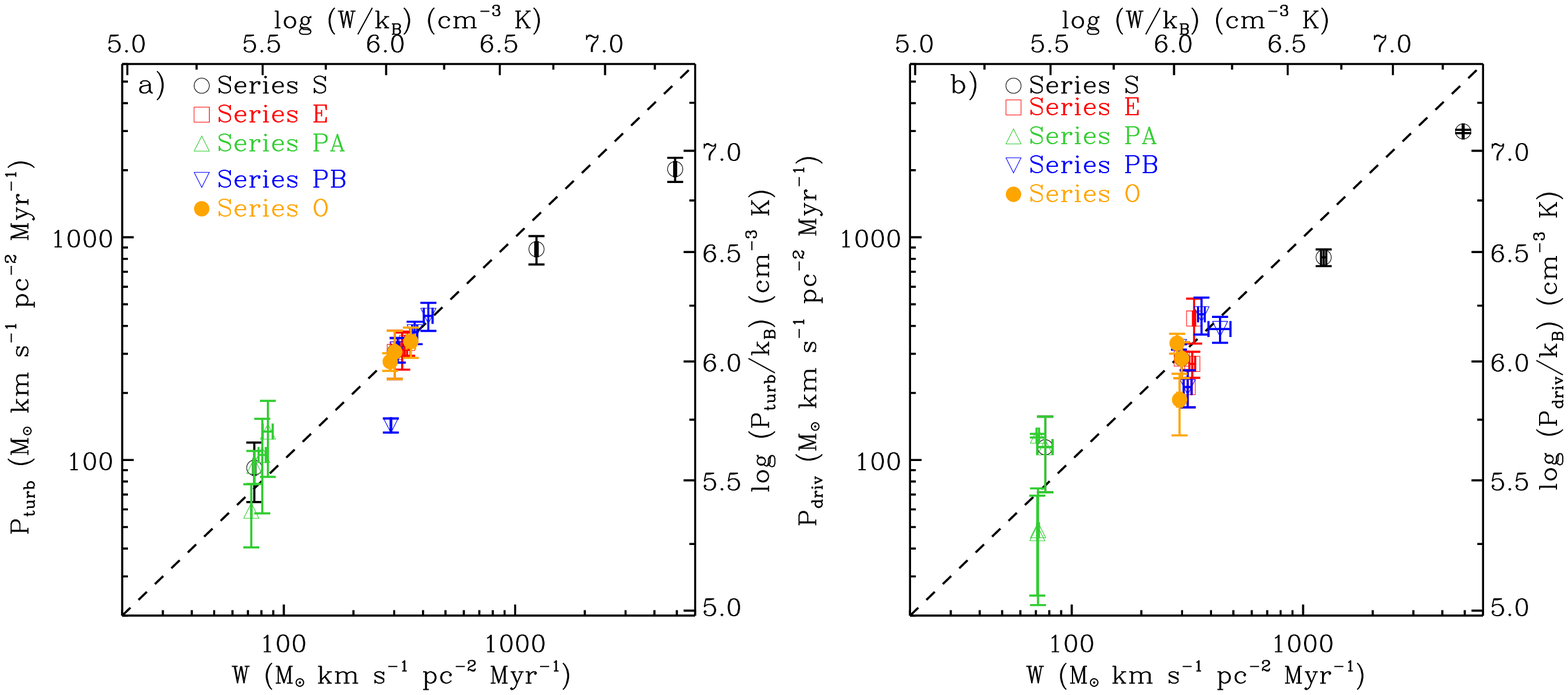}
\caption{Momentum fluxes (a) \pturb\ and (b) \pdriv\ plotted against
  vertical weight of gaseous disk $W$ from each simulation (see
  Equations \ref{wteqn} - \ref{pdrive}).  Each point indicates the mean
  values of the momentum fluxes measured in 20 Myr bins, after the
  simulations reach a steady state ($\approx$1 \torb).  The error bars
  show the (1$\sigma$) standard deviations.  The dashed
  lines show the expectation from vertical equilibrium with
  self-regulation, where \pturb\ $\approx W \approx$ \pdriv.}
\label{mfl}
\end{figure*}

Figure \ref{mfl} shows the relationships of the measured momentum
fluxes \pturb\ and \pdriv\ with $W$.  We compute $P_{\rm turb}$ in the
simulations using midplane horizontal- and time-averages of $\langle \rho
v_z^2\rangle$.  The turbulent and SN momentum fluxes are in excellent
agreement with the vertical weight of the disk.  For those models
showing the largest deviation from the expectations in Figure
(\ref{mfl}a), simulations PB1.5 and S800, there is only a factor of
two discrepancy between \pturb\ and $W$.  In Figure (\ref{mfl}b),
models S400 and S800 have the largest discrepancy between the
predicted and measured momentum injection rates.  For very strong
gravity models, the disk thickness becomes comparable to the (imposed)
radii of SN shells in our models.  As a consequence, the disk can
become ``artificially'' thickened, because real feedback shells
starting at much smaller radii and conserving momentum might not
expand as much.  If shells reach larger sizes than their ``natural''
radii, the corresponding mean density and pressure would be somewhat
lower than would be required for self-regulated equilibrium.  Overall,
the general correspondence between \pturb, \pdriv, and $W$ strongly
supports the idea that the evolution of our ISM models reaches an
equilibrium governed by star formation self-regulation.  To further
explore this premise, we now turn our attention to comparing other
physical properties of the simulations with the predictions from
self-regulation theory.

We begin by providing an overview of the analytical results expected
under self-regulation.  The star formation rate in equilibrium is obtained by
equating $f_p P_{\rm drive} \equiv P_{\rm turb}$ with $W$ (Equation
13 in Paper I):
\begin{eqnarray}
\Sigma_{\rm SFR} &=& \frac{2\pi(1+\chi)}{f_p} \frac{m_*G\Sigma^2}{p_*} 
\nonumber \\
&=& 0.092 \, {\rm M_\odot\, kpc^{-2} \, yr^{-1}}
\left(\frac{\Sigma}{100 \, {\rm M_\odot\,
    pc^{-2}}}\right)^2 \nonumber \\ 
& & \times \frac{(1+\chi)}{f_p} 
\left(\frac{p_*/m_*}{3000 \, {\rm \,km\, s^{-1}}}
\right)^{-1} .
\label{sigsfrpred}
\end{eqnarray}
The factor $\chi$ accounts for the gravitational potential due to the
bulge (see Section 2 and 4 in Paper I), with
\begin{equation}
\chi = \frac{2 C}{1+\sqrt{1+4C}}.
\label{Chi}
\end{equation}
Here, $C \approx 0.66 \mathcal{W}^2$, where
$\mathcal{W}=\sigma_{v}\Omega/(\pi G \Sigma)$ is a parameter analogous
to the Toomre $Q$ parameter \citep{Toomre64}.  Using typical values,
\begin{equation}
C = 0.35\left[ \left(\frac{\sigma_{v}}{{\rm 10 km \, s^{-1}}} \right)
\left(\frac{\Omega}{0.1 \, {\rm Myr}^{-1}}  \right)
\left( \frac{\Sigma}{100 \, {\rm M_\odot\, pc}^{-2}} \right )^{-1} \right ]^2;
\label{CW}
\end{equation}
thus $C$ is typically small in our simulations.

The parameter $f_p$ characterizes the magnitude of turbulent
dissipation, with $f_p \sim 1$ for strong dissipation and $f_p\sim 2$
for weak dissipation. The value of $f_p$ is defined by the ratio of
\pturb\ and the fiducial vertical momentum flux injected by star
formation, \pdriv\ (see Equations \ref{pturb} - \ref{pdrive}):
\begin{eqnarray}
\label{fpeqn1}
f_p &\equiv& P_{\rm turb} \left ( \frac{p_*}{4m_*}\Sigma_{\rm SFR}
\right )^{-1}\\
&=& W \left ( \frac{p_*}{4m_*}\Sigma_{\rm SFR}
\right )^{-1},
\label{fpeqn2}
\end{eqnarray}
where the second line assumes that dynamical equilibrium also holds
(see Fig. \ref{mfl}, as well as Fig. 8 of KKO11).  Accordingly, in a
self-regulated system,
\begin{eqnarray}
f_p &=& 0.92 \, (1+\chi) \, 
\left( \frac{\Sigma_{\rm SFR}}{\rm 0.1 M_\odot \, kpc^{-2} \, yr^{-1}
} \right)^{-1} 
 \left( \frac{\Sigma}{100 \, {\rm M_\odot\, pc^{-2}}}  \right )^2 \nonumber
\\
&& \times 
\left ( \frac{p_*/m_*}{3000 {\rm km \, s^{-1}}} \right )^{-1};
\label{fpexp}
\end{eqnarray}
this is a simply a re-arrangement of Equation (\ref{sigsfrpred}).  We
note that $P_{\rm turb}=\rho_0 \sigma_{v}^2$ is equivalent to $(\Sigma
\sigma_{v}/2)(H/\sigma_{v})^{-1}$.  With $\Sigma \sigma_{v}/2$ the
vertical momentum per unit area contained in each side of the disk,
and $H/\sigma_{v}$ the vertical crossing time, the relation $P_{\rm
  turb} \approx P_{\rm drive}$, or $f_p\approx 1$, thus implies that
the disk's vertical momentum is replenished by feedback approximately 
once per dynamical time.

Using the value of \sigsfr\ measured using Equation (\ref{csfr}) and
$\chi$ from Equation (\ref{Chi}), we can calculate $f_p$ in each
simulation from Equation (\ref{fpexp}).  To obtain $\chi$ through
Equations (\ref{Chi}) and (\ref{CW}), we use the
measured value of \vz.  Table \ref{paramtab} provides the values of
\fup\ measured from the simulations in this way; all values are near
unity.  We have verified that \fup\ measured through Equation
(\ref{fpeqn1}) provides similar values, since \pturb\ $\approx W$
(Fig. \ref{fpeqn2}).  Table \ref{paramtab} also shows that $\chi$ is
measured to be rather small.

By equating $P_{\rm turb}$ and $W$, the turbulent velocity dispersion
can be expressed as a relationship between the characteristic vertical
acceleration under self-gravity, $\sim \sigma_{v}/t_{\rm ff}$, and
the mean gravitational field $\approx \pi G\Sigma$:
\begin{eqnarray}
\sigma_{v}  &=& \frac{4}{\sqrt3} t_{\rm ff}(n_{\rm 0})  G \Sigma
(1+\chi)^{1/2}
\nonumber \\
 &=& 4.42 \, {\rm km \, s}^{-1}  \left ( \frac{n_0}{100 \, {\rm
    cm}^{-3}} 
\right )^{-1/2} \left ( \frac{\Sigma}{100 \, {\rm M_\odot\, pc}^{-2}}
\right )\nonumber \\ 
& & \times  (1+\chi)^{1/2}.
\label{vzpred2}
\end{eqnarray}
Equation (\ref{vzpred2}) should hold for any disk-like system 
supported primarily by
turbulence, independent of the source of that turbulence.

The predicted velocity dispersion \vz\ can also be expressed in terms of
\epsrhoo, $f_p$, and $\chi$ as: 
\begin{equation}
\sigma_{v}  = 5.5  \, {\rm km \, s^{-1}} \frac{f_p}{(1+\chi)^{1/2}}  \left
(  \frac{\epsilon_{\rm ff}(n_0)}{0.005} \right )
\left(\frac{p_*/m_*}{3000 \, {\rm \,km\, s^{-1}}} \right)
\label{vzpred1}
\end{equation}
(see Equation 22 of Paper I); Equation (\ref{vzpred1}) follows from 
Equation (\ref{vzpred2}) using the definitions of \epsrhoo\ and $f_p$
from Equations (\ref{epscalc}) and (\ref{fpexp}), respectively.
This form shows that if $f_p$ and \epsrhoo\ are approximately
constant, then the velocity dispersion would be proportional to the
momentum/mass injected by star formation.

Lastly, the predicted disk thickness \Ht\ when dynamical equilibrium
holds is:
\begin{eqnarray}
H &=& \frac{1}{1+\chi}\frac{\sigma_{v}^2}{\pi G \Sigma} \nonumber \\
&=& 74 \, {\rm pc} \frac{1}{1+\chi} \left(\frac{\sigma_{v}}{10 \,
{\rm \,km\, s^{-1}}}\right)^2 \left( \frac{\Sigma}{100 \, {\rm
M_\odot\, pc^{-2}}}\right)^{-1}.
\label{Hpred2}
\end{eqnarray}
(Note that the first equality in Equation 28 of Paper I contains a typo;
the denominator should contain a $\Sigma$ instead of a $\Sigma^2$.)
Using Equation (\ref{vzpred1}), this may be re-expressed as
\begin{eqnarray}
H &=& 23  \, {\rm pc} \, \frac{f_p^2}{(1+\chi)^2} \left (  \frac{\epsilon_{\rm ff}(n_0)}{0.005} \right )^2 \nonumber \\
& & \times \left ( \frac{p_*/m_*}{3000 \, {\rm \,km\, s^{-1}}} \right)^{2} \left(\frac{\Sigma}{100 \, {\rm M_\odot\, pc^{-2}}}\right)^{-1}. \nonumber \\
\label{Hpred}
\end{eqnarray}
When the definitions for \epsrhoo\ and $f_p$ (Equations \ref{epscalc} and
\ref{fpexp}) are substituted into Equation (\ref{Hpred}), the result
is $\Sigma/(2\rho_0)$.  While Equation (\ref{Hpred2}) should hold
independent of the source of turbulence, Equation (\ref{Hpred}) shows
that \Ht\ would scale inversely with \sig\ for self-regulated
turbulent disks if $f_p$ and \epsrhoo\ remain approximately constant.

Using the measured values of \sigsfr, \vz, \Ht, and \no\ in each
simulation, we can test a number of aspects of the theory in Paper I.
In particular, we can: (1) compare our measurements of \sigsfr\ to
Equation (\ref{sigsfrpred}) to assess the combined (turbulent
driving/dissipation and gravity/pressure) equilibrium and test whether
$f_p\sim 1$ is satisfied (for varying physical parameters \sig, \psn,
\omg\ and varying numerical parameter \epsnth); (2) compare our
measurements of \vz\ to Equation (\ref{vzpred2}) to assess the balance
of turbulent pressure and weight, also comparing to Equation
(\ref{vzpred1}) to evaluate whether $f_p$ and \epsrhoo\ are
effectively constant (for varying parameters); (3) compare our
measurements of \Ht\ to Equation (\ref{Hpred2}) to assess dynamical
equilibrium, also comparing to Equation (\ref{Hpred}) to evaluate
whether $f_p$ and \epsrhoo\ are effectively constant (for varying
parameters).  In addition, we can (4) use our measurements of
\sigsfr\ and \no\ to compute a measured \epsrhoo\ via Equation
(\ref{epscalc}) and explore whether there are any systematic
dependencies on the physical or numerical parameters.

\begin{figure*}
  \includegraphics[width=15cm]{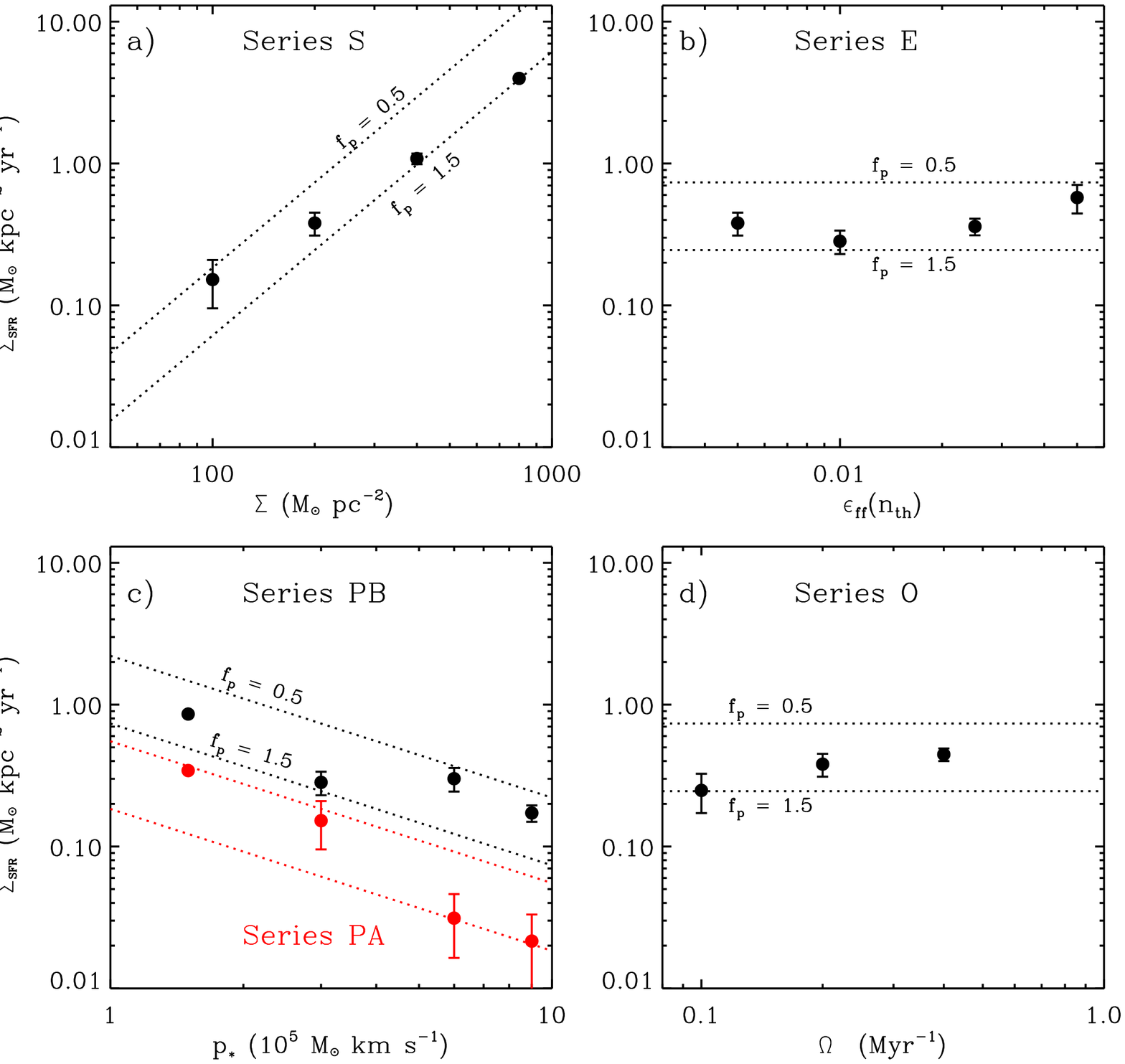}
\caption{Mean \sigsfr\ from the simulations (points), and that
  predicted from self-regulated equilibrium (lines), for two different
  values of $f_p$ = 0.5 and 1.5 (see
  Equation \ref{sigsfrpred}). \sigsfr\ is plotted against the model input
  parameter that is varied in each series: a) gas surface density
  $\Sigma$, b) star formation efficiency in dense gas \epsnth, c)
  momentum input per high mass star \psn, and d) local angular
  rotational velocity $\Omega$ (which also affects the vertical
  gravity).  Error bars show the 1$\sigma$ deviation of the measured
  \sigsfr.}
\label{sfrpred_all}
\end{figure*}

Figure \ref{sfrpred_all} shows the mean \sigsfr\ for all models after
a steady state is reached (generally $t > 50$ Myr).  The star
formation rate is plotted against the main user-defined parameters
varied between models from each series, a) \sig\ (Series S), b)
\epsnth\ (Series E), c) \psn\ (Series PA and PB), and d) $\Omega$
(Series O).  The dashed lines in each panel show the predictions from
self-regulation theory (Equation \ref{sigsfrpred}), for $f_p$ = 0.5
and 1.5, and with $\chi$=0.

For Series S, Figure \ref{sfrpred_all}a shows a remarkably good
agreement between the measured star formation rate and the prediction
for $f_p\sim 1$.  Although the increase of \sigsfr\ with \sig\ for
Series S is slightly shallower than the power predicted in Equation
(\ref{sigsfrpred}) (1.6 vs. 2), a larger adopted \epsnth\ leads to a
slightly steeper slope, so that our overall results are generally
consistent with $\Sigma_{\rm SFR} \propto \Sigma^2$ (see Fig. 4 of
Paper I).

Equation (\ref{sigsfrpred}) indicates that \sigsfr\ under
self-regulation is independent of the star formation efficiency in
dense gas.  Figure \ref{sfrpred_all}b indeed shows that the measured
value of \sigsfr\ for Series E models is relatively insensitive to the
chosen value of \epsnth.  Physically, this means that (within limits)
the rate of star formation in dense gas does not affect the overall
star formation rate averaged over large scales, because the amount of
mass at high density simply adjusts until the feedback rate matches
what is required to produce the needed turbulent pressure.

From Equation (\ref{sigsfrpred}), the star formation rate in
self-regulated equilibrium should be inversely proportional to the
input momentum per stellar mass $p_*/m_*$, where $p_*$ is associated
with high-mass stars and $m_*$ includes all of the lower-mass stars
proportionally (based on the IMF).  Figure \ref{sfrpred_all}c shows
\sigsfr\ as a function of \psn\ for Series PA and PB models.  An
inverse proportionality between \sigsfr\ and \psn\ is evident,
comparing favorably to the prediction from self-regulation.

The rate of star formation in a self-regulated system is not expected
to depend on the angular velocity \omg, provided that angular momentum
does not limit local collapse (i.e. on scales \aplt $H$) 
and that the vertical stellar gravity is small
compared to the vertical gas gravity (i.e. $\chi \ll 1$).  The
independence of \sigsfr\ from \omg\ is shown in Figure
\ref{sfrpred_all}d.  We found, however, that when \omg\ is large
enough -- as in Model O8, angular momentum prevents clouds from
collapsing to reach high densities (\nth\ = 5000 \cmt), and we
register it as non-star-forming.  This model has Toomre wavelength
$\lambda_T = \pi^2 G \Sigma/\Omega^2 \approx $ 14 pc.  This value of
$\lambda_T$ is comparable to what would otherwise be the collapse
scale, thereby stabilizing the ISM and preventing the formation of any
clouds.

\begin{figure}
  \includegraphics[width=8cm]{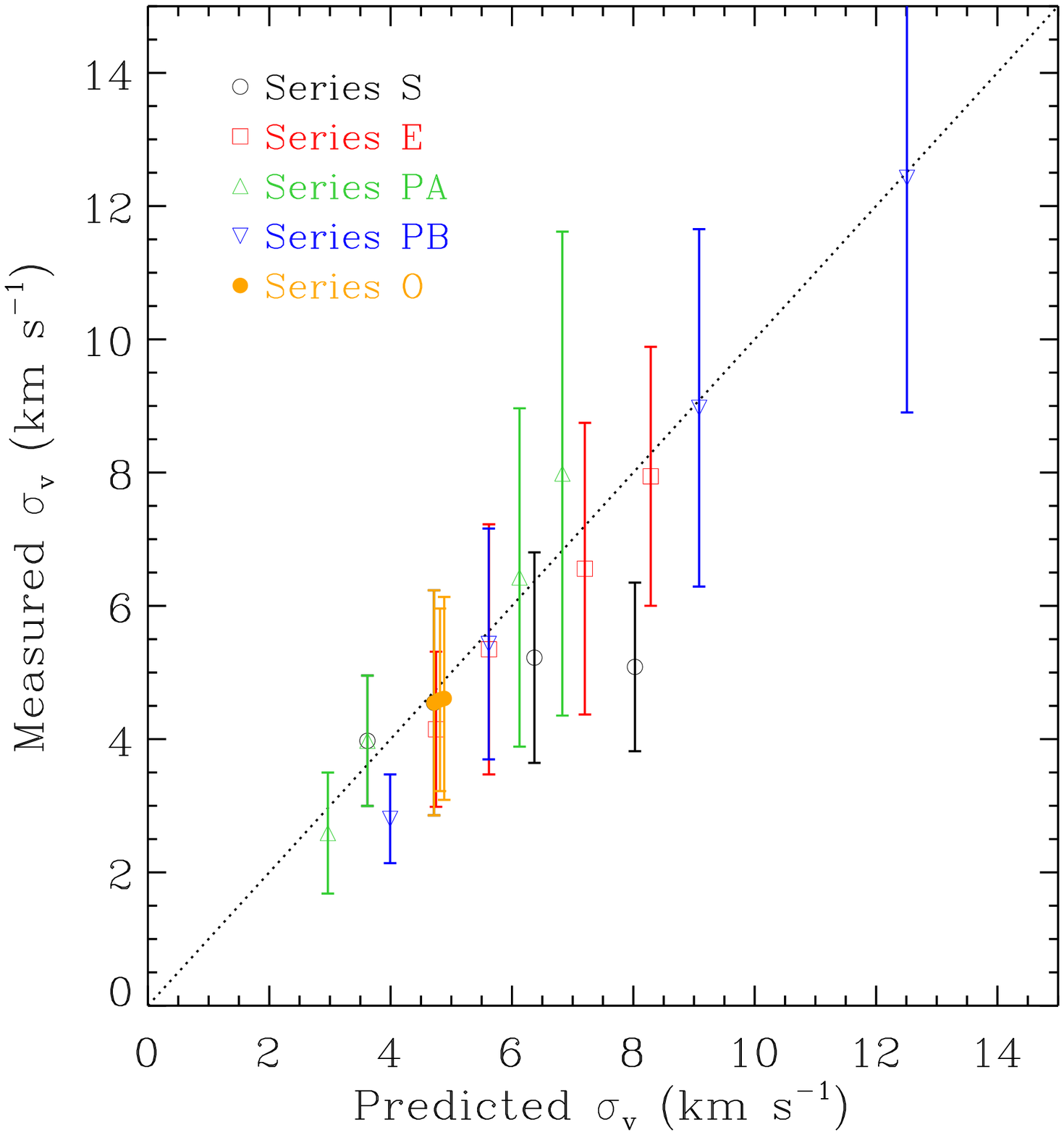}
\caption{The measured and predicted values of the velocity dispersion
  \vz\ for all models, as defined in Equations (\ref{wvz}) and
  (\ref{vzpred2}), respectively. The agreement between the measured
  and predicted dispersion shows that dynamical equilibrium between
  gravity and turbulent pressure is established.}
\label{compvz}
\end{figure}

Figure \ref{compvz} shows how \vz\ as measured in each simulation
(Equation \ref{wvz}) compares to the expectation from vertical
dynamical equilibrium (Equation \ref{vzpred2}).  There is generally a
good correspondence between the predicted and measured values.  This
comparison contains essentially the same information as in Figure
(\ref{mfl}a), and similar to the results there, the measured \vz\ for
a few models depart somewhat from the prediction.  The greatest
departure is for model S800, which is expected since the disk
thickness approaches the numerically-imposed feedback shell size.

\begin{figure*}
  \includegraphics[width=15cm]{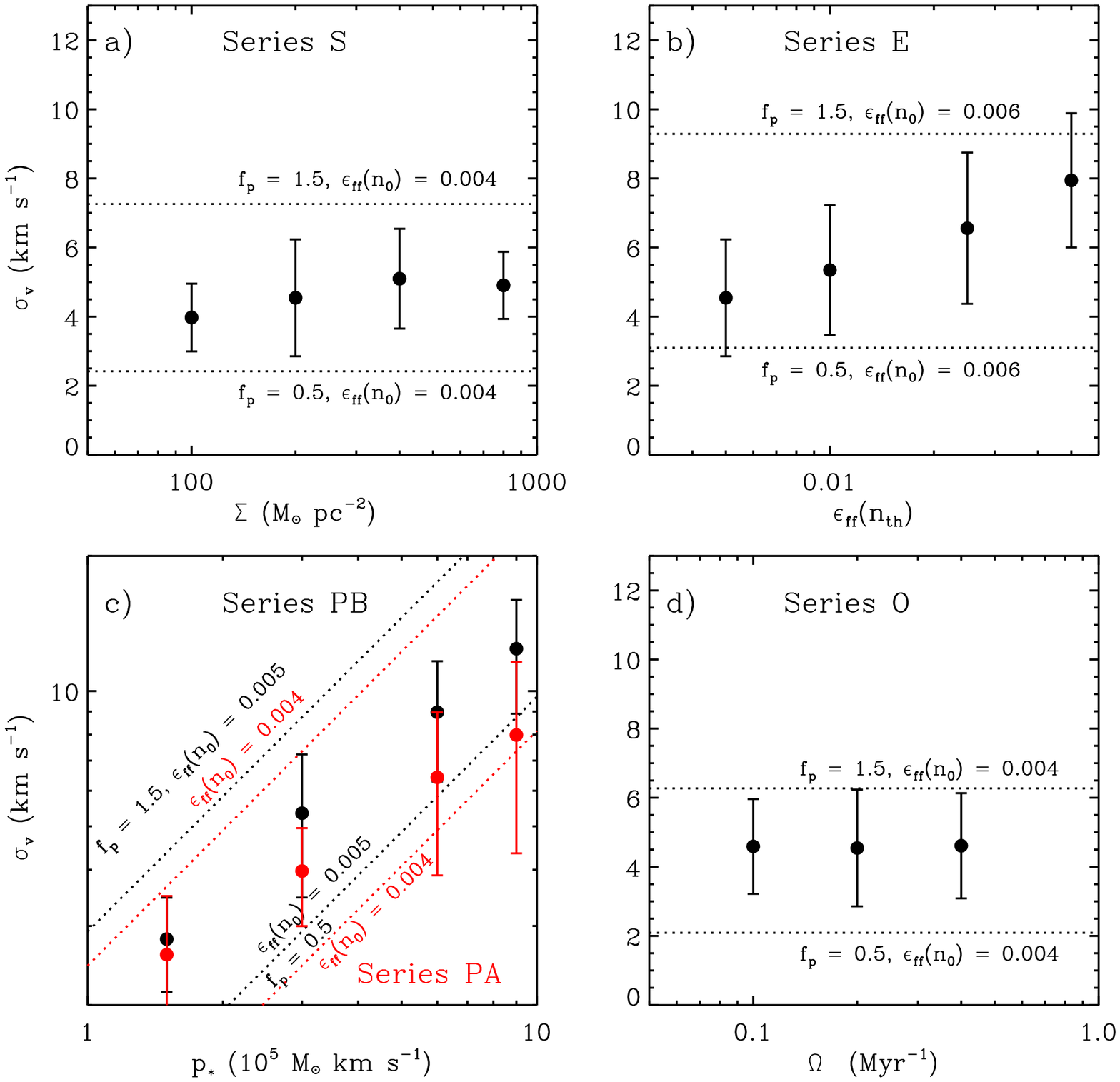}
\caption{Mean vertical velocity dispersion \vz\ (see Equation
  \ref{wvz}) from simulations (points), along with the prediction from
  self-regulation (lines), assuming the mean value of \epsrhoo\ for
  each Series and with $f_p$=0.5 and 1.5 (see Equation \ref{vzpred1}).
  The panels separately show each series, i.e.  \vz\ against a)
  $\Sigma$, b) \epsnth\, c) \psn, and d) $\Omega$.  Error bars show
  the 1$\sigma$ deviation of the measured \vz.}
\label{preddisp}
\end{figure*}

Figure \ref{preddisp} shows \vz\ measured in the simulations for each
series.  The dashed lines in each panel indicate the prediction from
self-regulation given by Equation (\ref{vzpred1}), again with
$f_p=$0.5 and 1.5, along with $\chi$=0, and using the mean value of
\epsrhoo\ for each series.  The predicted independence of \vz\ from
\sig\ and \omg\ is confirmed in Figures \ref{preddisp}a and
\ref{preddisp}d, respectively.

Figures \ref{preddisp}b-c indicate that the measured \vz\ increases
with \epsnth\ and \psn, respectively.  In Figure \ref{preddisp}b, the
lines correspond to Equation (\ref{vzpred1}) with constant
\epsrhoo=0.006.  However, as discussed below, the measured
\epsrhoo\ increases with \epsnth\ (also evident in Table
\ref{paramtab}), implying from Equation (\ref{vzpred1}) that
\vz\ should indeed increase with \epsnth.  Similarly, Figure
\ref{preddisp}c shows that the increase in \vz\ with \psn\ is
shallower than the linear relation indicated in Equation
(\ref{vzpred1}) for constant \epsrhoo.  This is due to the slight
decrease in \epsrhoo\ with \psn, which is further discussed below.

\begin{figure}
  \includegraphics[width=8cm]{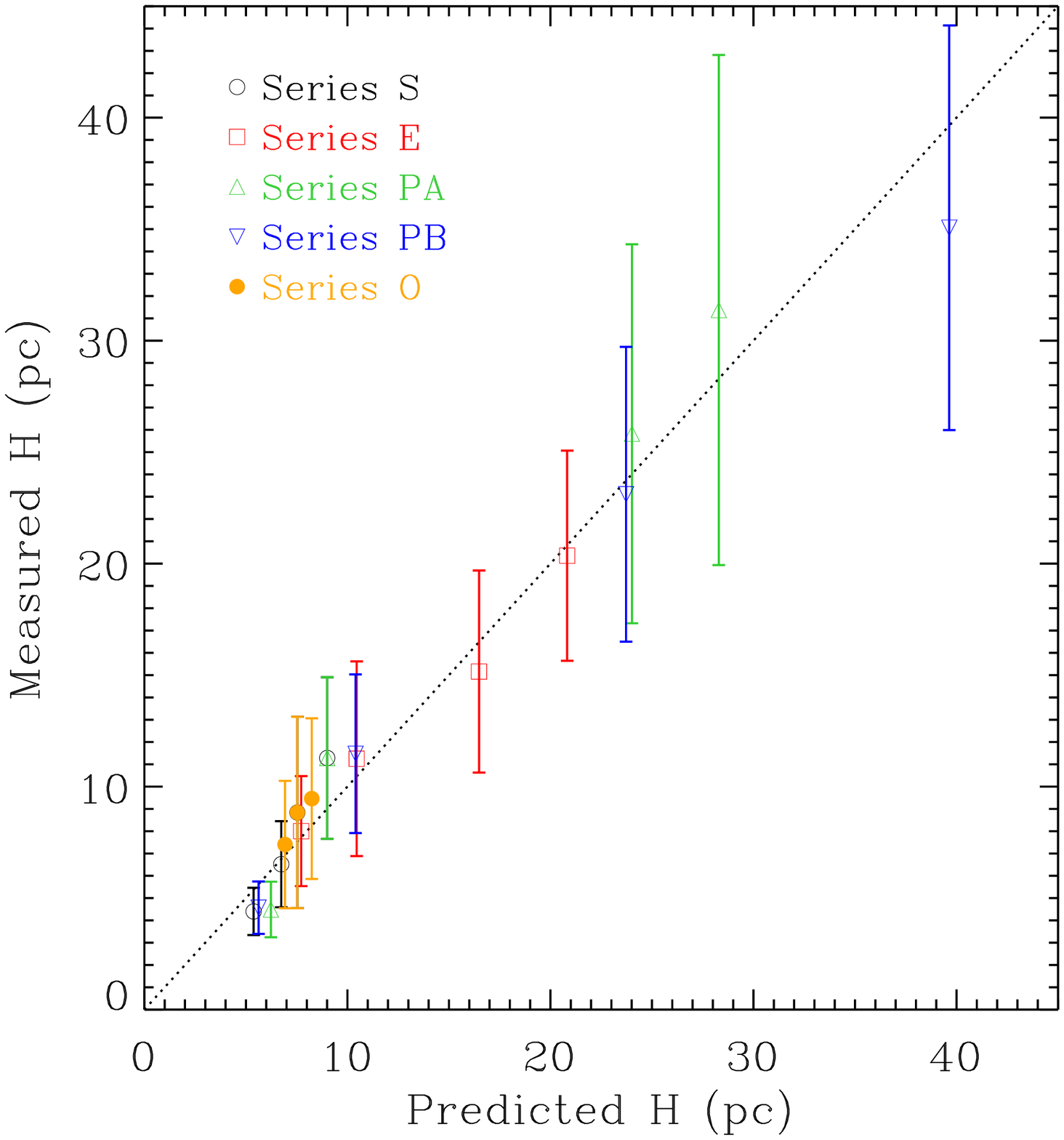}
\caption{The measured and predicted values of the disk thickness
  \Ht\ for all models, as defined in Equations (\ref{wh}) and
  (\ref{Hpred2}), respectively.  The agreement between the
  measured and predicted thickness indicates vertical equilibrium
  between the weight due to gravity and SN driven turbulent pressure.}
\label{comph}
\end{figure}

Figure \ref{comph} compares the measured and predicted values of \Ht,
given by Equation (\ref{wh}) and (\ref{Hpred2}) respectively.  As with
the velocity dispersion (Fig. \ref{compvz}), the thickness is measured
to be very similar to the predicted value.  This agreement is
indicative of vertical equilibrium between the weight due to gravity
and turbulent pressure.

\begin{figure*}
  \includegraphics[width=15cm]{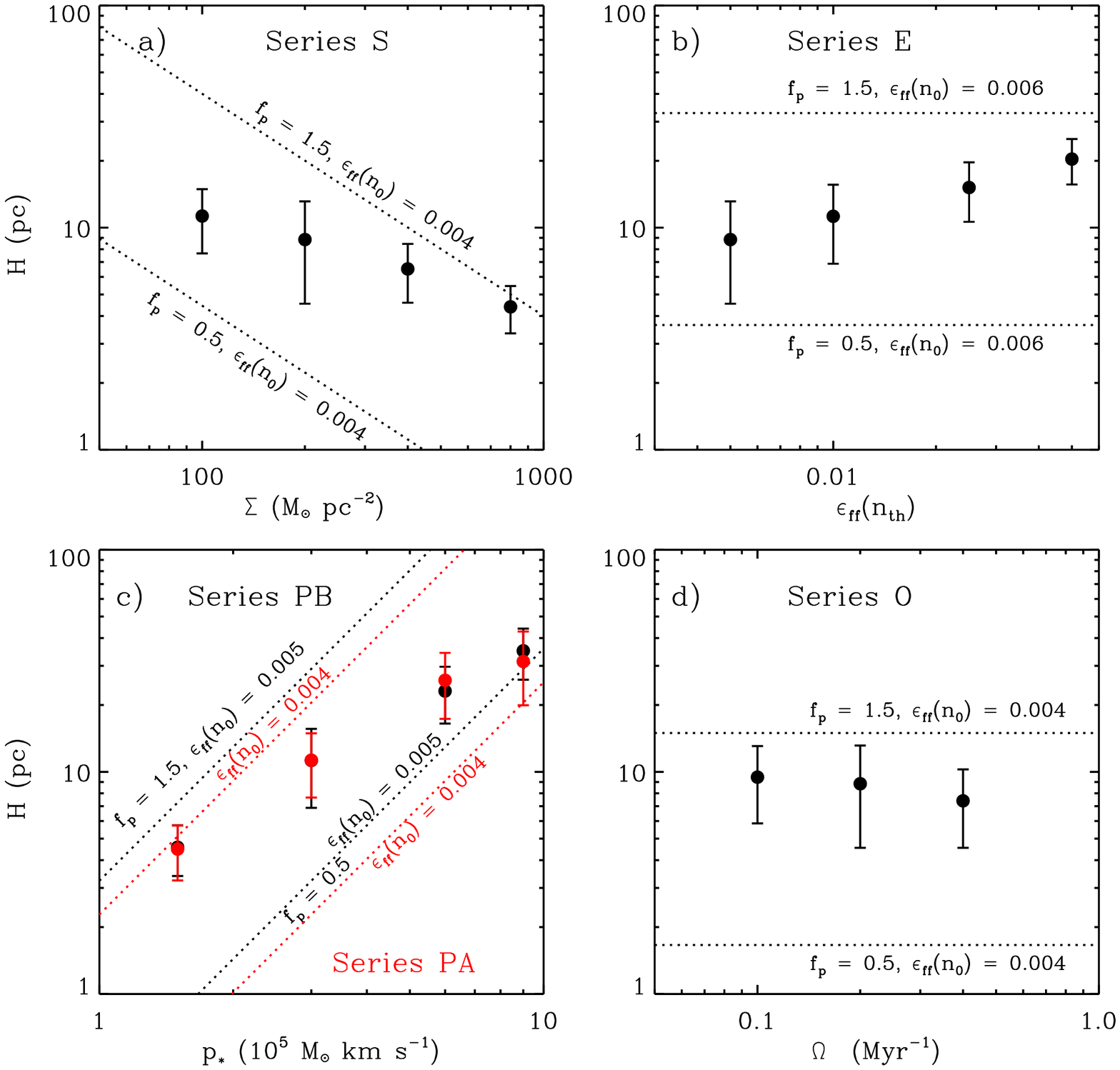}
\caption{Mean disk thickness $H$ (Equation \ref{wh}) in all models,
  along with the prediction from self-regulation (lines), assuming the
  mean value of \epsrhoo\ for each Series and with $f_p$=0.5 and 1.5
  (see Equation \ref{Hpred}).  $H$ is plotted against a) $\Sigma$, b)
  \epsnth\, c) \psn, and d) $\Omega$.  Error bars show the 1$\sigma$
  deviation of the measured \Ht.}
\label{Hpredfig}
\end{figure*}

Figure \ref{Hpredfig} shows the measured thickness \Ht\ from Equation
(\ref{wh}), compared to the prediction from Equation (\ref{Hpred}) for
constant $f_p$ and \epsrhoo.  Figures \ref{Hpredfig}a shows that
\Ht\ decreases with \sig\ less steeply than $\propto
\Sigma^{-1}$. Based on Equation (\ref{Hpred}) this is consistent with
the systematic increase of $f_p$ with \sig\ for Series S (see Table
\ref{paramtab}).  Since the SN shell radius is chosen to be 5 pc in
these numerical simulations, this places an effective lower limit on
the disk thickness \Ht\ \apgt 5 pc, which might be part of the reason
for the shallow decrease of $H$ with $\Sigma$.  Similar to the case of
\vz, the increase of \Ht\ with \epsnth\ and \psn\ in Figures
\ref{Hpredfig}b-c can be fully accounted for by the variation of
\epsrhoo\ with \epsnth\ and \psn, respectively.  Figure
\ref{Hpredfig}d shows that \Ht\ is insensitive to \omg, implying that
neither rotation nor the external gravity of the bulge strongly
affects the disk thickness, within the range shown.

\begin{figure*}
  \includegraphics[width=15cm]{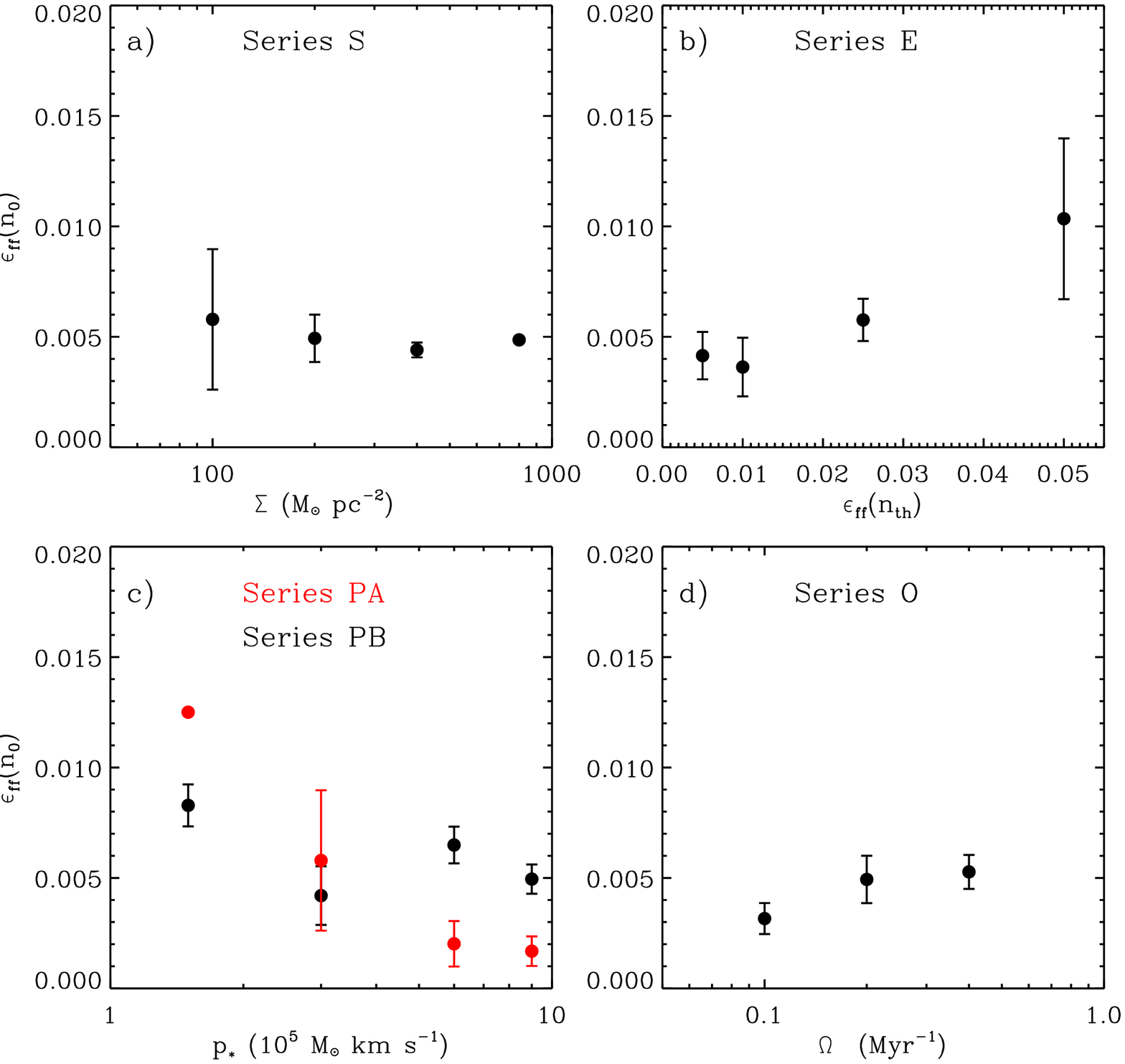}
\caption{Calculated star formation efficiency per free-fall time at
  the midplane density, \epsrhoo, plotted for all models  
  against a) $\Sigma$, b) \epsnth\, c) \psn, and d) $\Omega$.  Error
  bars show the 1$\sigma$ deviation of the measured \epsrhoo.}
\label{epspredfig}
\end{figure*}

The measured value of \epsrhoo, as computed through Equation
(\ref{epscalc}) for each simulation, is shown in Figure
\ref{epspredfig}.  In general, there are only slight variations in
\epsrhoo\ among the simulations; for Series S and O, \epsrhoo\ is
approximately constant.  As discussed above with regard to \sigsfr, we
interpret the weak dependence of \epsrhoo\ on \epsnth\ as indicative
of an adjustment in the mass of dense gas to meet the large-scale need
for star formation feedback.  This adjustment is possible because the
dynamical timescales decrease with increasing density and decreasing
spatial scale.  Other recent numerical studies have also found that
large-scale star formation rates are insensitive to user-defined
parameters controlling star formation at small scales (see section
\ref{prevwork}).  Figure \ref{epspredfig}c demonstrates that
\epsrhoo\ decreases somewhat with increasing \psn.  Potentially, this
may be due to the increase of velocity dispersion with increasing
\psn, which renders a smaller fraction of gas eligible to collapse
(see section \ref{prevwork}).

In summary, based on our quantitative comparisons, the results from
our numerical simulations show good agreement with the simple analytic
theory of Paper I.  Both vertical dynamical equilibrium and a balance
between turbulent driving and dissipation are satisfied.  The
dependence of \sigsfr, \vz, and \Ht\ on the gas surface density \sig\
and input momentum \psn\ are similar to the predicted behavior.  In
addition, the results are insensitive to the exact prescription for
star formation in dense gas.  The free parameter $f_p$
was introduced in Paper I to characterize the turbulent ``yield'' from
momentum inputs by star formation, and our present simulations provide
a numerical evaluation of $f_p$.  For our whole simulation suite, $f_p$
remains within $\sim 50$\% of unity, the value for strong dissipation.

\section{Discussion and Summary}\label{discsec}

To investigate dynamics of the highly-turbulent, molecule-dominated
ISM as found in (U)LIRGS and galactic centers, we have executed a suite
of numerical simulations that incorporate feedback from star
formation.  We demonstrate that in simulations reaching a steady
state, many physical properties can be accounted for by a simple
theory of star formation self-regulation (Paper I).  Namely, the turbulent
pressure is driven by injected SN momentum, and dissipates within a
vertical crossing time of the disk.  The rate of star formation and
momentum injection adjusts until the input rate of momentum flux
balances the vertical weight of the gaseous disk.

\subsection{Relationship to Previous Work}\label{prevwork}

Our numerical simulations of the ISM are similar in some respects to
other recent modeling efforts that have included turbulent driving
from localized feedback events, and our results are consistent with
previous findings.  In particular, we have found that the velocity
dispersion \vz\ is not strongly dependent on the the exact
prescription for feedback as long as the momentum (or energy) input is
similar \citep[][KKO11]{Dib+06, Shetty&Ostriker08, Joung+09}.
Additionally, we find that the overall star formation rate \sigsfr\ is
not sensitive to the chosen value of \epsnth, in general agreement
with the conclusions of \citet{Dobbs+11b} and \citet{Hopkins+11} that
the specific small-scale star formation prescription does not not
strongly affect the resulting \sigsfr.  Similar to previous efforts
that have explored a large range of surface densities, our simulations
here and in Paper I 
clearly demonstrate a power law relationship between \sig\ and
\sigsfr\ \citep[e.g.][]{Lietal05,Lietal06,Tasker&Bryan06,Tasker&Bryan08,
  Robertson&Kravtsov08,Shetty&Ostriker08, Dobbs&Pringle09,
  KoyamaOstriker09a, Dobbs+11b, Hopkins+11}.  Here, our numerical
simulations are well resolved in the vertical direction, and we relate
both the power law and coefficient of the
\sigsfr\ vs. \sig\ relationship to the requirements for equilibrium
given in the self-regulation theory of Paper I.

For the high-surface-density regime \sig\apgt 100 \msunpc\ studied in
this work, observations show that the \sigsfr\ vs. \sig\ relationship
is steeper than linear.  As discussed in Paper I, an accurate
calibration of \XCO, the ratio of gas mass to velocity-integrated CO
intensity, is crucial for estimating \sig\ (and thus the exact power
law of \sigsfr\ vs. \sig) from CO observations.  Recent theoretical
efforts have advanced our understanding of \XCO, in both Milky-Way
like GMCs \citep[e.g.][]{Glover&MacLow11,Shetty+11a, Shetty+11b}, as
well as large scale galaxies and merger systems
\citep[e.g.][]{Narayanan+11a,Narayanan+12}.  These models use a
combination of numerical hydrodynamic simulations and radiation
transfer to assess environmental dependencies of \XCO.

If gas dominates the vertical gravity, the theory of Paper I results
in a power-law relationship \sigsfr $\propto$ \sig$^2$ (Equation
\ref{sigsfrpred}); the numerical simulations presented in Paper I and
here (Fig. \ref{sfrpred_all}) support this model.  As demonstrated in
Paper I, employing a continuously varying \XCO\ indeed shows \sigsfr
$\propto$ \sig$^2$ for a sample of ULIRGs and the Galactic center
\citep{Genzel+10,Yusef-Zadeh+09}.  \citet{Narayanan+12} investigated
the relationship between $X_{\rm CO}$ and the velocity integrated CO
($J=1-0$) brightness temperature $W_{\rm CO}$ in a large compilation of
low- and high-$z$ galaxies.  Applying the model-based calibration
$X_{\rm CO} \propto W_{\rm CO}^{-0.3}$, \citet{Narayanan+12} found
that $\Sigma_{\rm SFR} \approx 0.1 \, {\rm M_\odot\, kpc^{-2} \,
  yr^{-1}} ({\Sigma}/{100 \, {\rm M_\odot\,pc^{-2}}})^{1.95}$, in
agreement with the Paper I prediction (Equation \ref{sigsfrpred} here,
with $f_p\approx 1$ and $\chi \ll 1$).

The self-regulation theory of Paper I has a number of similarities to
and differences from the star formation model in the
high-surface-density molecule-dominated regime (\sig\apgt 100 \msunpc)
proposed by Krumholz and coworkers \citep{Krumholz&McKee05,
  Krumholz+09, Krumholz+12}.  Both models rely on the role of
supersonic turbulence.  In Krumholz et al, the specific star formation
rate is characterized in terms of an efficiency per free-fall time at
the mean density (essentially \epsrhoo), where the mean density (which
sets \tffno) depends on \sig\ and the turbulence level.
\citet{Krumholz&McKee05} argued that this efficiency should depend on
the fraction of gas at pressures higher than the mean turbulent
pressure, and pointed out that for log-normal density PDFs, this
fraction depends only weakly on the Mach number ($\propto {\cal
  M}^{-0.3}$) and is predicted to be $\sim 0.01$, consistent with
observations of molecular gas \citep{Krumholz&Tan07}.  Krumholz et al.
do not, however, directly address the origin of turbulence --
i.e. what sets \vz.  Rather, they adopt the assumption that Toomre $Q$
(and therefore $\cal W$) is order-unity, so that $\sigma_{v} \sim \pi
G \Sigma/\Omega$, and adopt an empirically-motivated relation $\Omega \propto
\Sigma^{0.5}$ (so that $\sigma_v \propto \Sigma^{0.5}$) 
to obtain $\Sigma_{\rm SFR} \propto \epsilon_{\rm ff}
\Sigma/t_{\rm ff} \propto \Sigma^2/\sigma_{v}^{1.3} \propto
\Sigma^{1.3} $.

Although the star formation rate in the current theory can also be
characterized in terms of the velocity dispersion and \epsrhoo\ (see
Equation 21 of Paper I), the fundamental relationship is instead Equation
(\ref{sigsfrpred}).  This expression connects the star formation rate
to the weight of the ISM (or equilibrium midplane pressure) and to the
momentum/mass injected by feedback ($p_*/m_*$), yielding $\Sigma_{\rm
  SFR} \approx 2 \pi G \Sigma^2 (p_*/m_*)^{-1}$.  Equating this
relation to \sigsfr $\equiv$ \epsrhoo \sig/\tffno $\approx (4
\epsilon_{\rm ff}(n_0)/\sqrt{3})G\Sigma^2/\sigma_{v}$ then leads to a
proportionality between the velocity dispersion and both \epsrhoo\ and
$p_*/m_*$ (see Equation \ref{vzpred1}).  Here, we use numerical
simulations to evaluate \epsrhoo\ (see Fig. \ref{epspredfig} and Table
\ref{paramtab}), finding values in the range $\sim 0.005-0.01$,
consistent with \citet{Krumholz&McKee05} and \citet{Krumholz&Tan07}.
We find, however, that the velocity dispersion is essentially
independent of \sig\ (see Fig. \ref{preddisp}a), which differs from
the $\sigma_{v} \propto \Sigma^{0.5}$ relation adopted by Krumholz et
al.  We note that due to lack of
resolution, the molecular velocity dispersion on scales comparable to
the disk thickness is difficult to obtain with observations, although
this situation will improve with ALMA.

\subsection{Summary of Results}\label{sumsec}

We have conducted a suite of simulations in which we independently
varied the gas surface density \sig, the input momentum per high mass
star $p_*$, the angular rotation rate of the gas \omg, and the
efficiency of star formation in very dense gas, \epsnth.  For each
simulation, we measured the star formation rate and ISM properties
after a statistical steady state developed, and compared to the
predictions of Paper I.  Our main results are as follows:

[1] For essentially all models, we find excellent correspondence
between the turbulent momentum flux \pturb, the vertical weight of the
gaseous disk $W$, and the vertical momentum injection rate per area
\pdriv\ associated with feedback (Fig. \ref{mfl}).  The result that
\pturb\ $\approx W \approx$ \pdriv\ strongly supports the idea that
the combined ISM/star formation system in starburst disks can be
self-regulated, as described in Paper I.

[2] Our results (Fig. \ref{sfrpred_all}) show that \sigsfr\ is
essentially independent of \omg\ and \epsnth, whereas
\sigsfr\ increases for higher \sig\ and decreases for higher $p_*$
following the expectations of self-regulation theory.  The result that
the large-scale \sigsfr\ is independent of the star formation rate in
dense gas means that the processes with the longest timescales
(associated with the largest spatial scales) are what controls the
overall star formation rate.  Physically this makes sense: gas that
reaches high density collapses rapidly, but the (slower) rate at which this
dense gas is resupplied by lower-density gas depends on larger-scale
dynamics.  As noted in Paper I \citep[see also ][]{Narayanan+12}, the
prediction $\Sigma_{\rm SFR} \approx 0.1 \, {\rm M_\odot\, kpc^{-2} \,
  yr^{-1}} ({\Sigma}/{100 \, {\rm M_\odot\,pc^{-2}}})^2$ of Equation
(\ref{sigsfrpred}) also agrees with observations provided that
\XCO\ decreases modestly with increasing \sig\ (or $W_{\rm CO}$).

[3] We find that the star formation efficiency per free-fall time at
the mean midplane density, \epsrhoo, is independent of \sig, \omg, and
\epsnth, and decreases only slightly with increasing $p_*$.  The
resulting \epsrhoo $\sim 0.005-0.01$ is similar to the theoretical
estimates for turbulent, self-gravitating gas at high Mach number of
\citet{Krumholz&McKee05}, while being somewhat lower than the
numerical estimates (from turbulent simulations with periodic boundary
conditions) of \citet{Padoan&Nordlund11b}.  Measured ratios of the
stellar-to-gas content in nearby molecular clouds are $\sim 0.03-0.06$
\citep{Evans+09}, which would imply similar \epsrhoo\ to our results
if the cloud ages are several free-fall times.  Gas at more extreme
conditions in ULIRGs is also observed to have \epsrhoo $\sim 0.01$
\citep{Krumholz&Tan07}.

[4] The vertical velocity dispersions in our models are in the range
$\sigma_{v}\in 3-12$ \kms\ for momentum per high mass star in the
range $p_*\in1.5-9 \times 10^5$ \msun\, \kms.  Similar to previous
results, we find that \vz\ is relatively independent of \sigsfr\ and
also \omg\ (see Fig. \ref{preddisp}).  The increase of $v_z$ with
$p_*$ is shallower than linear, due to the decrease of \epsrhoo\ with
increasing $p_*$ (see Equation \ref{vzpred1}).  The agreement (Figs.
\ref{compvz} and \ref{comph}) between \vz\ and \Ht\ measured in the
simulations and the respective predictions of Equations
(\ref{vzpred2}) and (\ref{Hpred2}) shows that dynamical equilibrium
between gravity and turbulent pressure is established.  The disk
thicknesses in our models are quite low ($H \sim 4-40\,{\rm pc}$),
increasing as $p_*$ increases and decreasing as \sig\ decreases.

[5] The densities and velocities in our models follow approximately
log-normal and normal distributions, respectively (Figs. \ref{denpdf},
\ref{velpdf}).  These forms are a natural consequence of supersonic
isothermal turbulent flows \citep[as discussed
  by][]{Vazquez-Semadeni94,Klessen00,Ostriker+99,Ostriker+01}.  The
log-normal density distribution is a key feature invoked in various
models of what sets the star formation efficiency in turbulent systems
\citep{Krumholz&McKee05, Padoan&Nordlund11b, Hennebelle&Chabrier11}.

A natural extension of the simulations presented here is to include
the third dimension. High resolution 3D simulations will allow for
detailed morphological and kinematic studies of the molecular ISM in starburst
regions.  Such simulations will also more accurately measure the
parameter $f_p$ relating the turbulent pressure to the momentum flux
injected by feedback (Equation \ref{fpeqn1}).  Further, more realistic
modeling of the ISM should incorporate a variety of feedback mechanisms
and additional physics, including stellar winds and radiation,
and heating and cooling to follow cold, warm, and hot phases rather than an 
isothermal equation of state to follow just the cold gas.  
By combining with radiative
transfer calculations, such simulations will enable detailed comparison of 
feedback-regulated disks with observations of the ISM
in starburst environments.

\acknowledgements

We are grateful to Frank Bigiel, Alberto Bolatto, Michael Burton, Paul
Clark, Roland Crocker, Simon Glover, Chang-Goo Kim, Woong-Tae Kim,
Ralf Klessen, Chris McKee, Desika Narayanan, and Rowan Smith for
useful discussions regarding star formation and molecular gas, and to
the referee for helpful comments on the manuscript.  The simulations
presented here were performed on the Odyssey cluster, supported by the
Harvard FAS Research Computing Group, and the Deepthought cluster
operated by the Astronomy CTC at the University of Maryland.  RS is
supported by the German Bundesministerium f\"ur Bildung und Forschung
via the ASTRONET project STAR FORMAT (grant 05A09VHA), as well as the
Deutsche Forschungsgemeinschaft (DFG) via the SFB 881 (B1 and B2)
``The Milky Way System'' and the SPP (priority program) 1573.  The
research of ECO is supported by grant AST-0908185 from the National
Science Foundation.

\bibliography{citations}

\begin{thebibliography}{81}
\expandafter\ifx\csname natexlab\endcsname\relax\def\natexlab#1{#1}\fi

\bibitem[{{Bally} {et~al.}(1987){Bally}, {Stark}, {Wilson}, \&
  {Henkel}}]{Bally+87}
{Bally}, J., {Stark}, A.~A., {Wilson}, R.~W., \& {Henkel}, C. 1987, \apjs, 65,
  13

\bibitem[{{Bally} {et~al.}(1988){Bally}, {Stark}, {Wilson}, \&
  {Henkel}}]{Bally+88}
---. 1988, \apj, 324, 223

\bibitem[{{Bigiel} {et~al.}(2010){Bigiel}, {Leroy}, {Walter}, {Blitz},
  {Brinks}, {de Blok}, \& {Madore}}]{Bigiel+10}
{Bigiel}, F., {Leroy}, A., {Walter}, F., {Blitz}, L., {Brinks}, E., {de Blok},
  W.~J.~G., \& {Madore}, B. 2010, \aj, 140, 1194

\bibitem[{{Bigiel} {et~al.}(2008){Bigiel}, {Leroy}, {Walter}, {Brinks}, {de
  Blok}, {Madore}, \& {Thornley}}]{Bigiel+08}
{Bigiel}, F., {Leroy}, A., {Walter}, F., {Brinks}, E., {de Blok}, W.~J.~G.,
  {Madore}, B., \& {Thornley}, M.~D. 2008, \aj, 136, 2846

\bibitem[{{Blitz} \& {Rosolowsky}(2004)}]{Blitz&Rosolowsky04}
{Blitz}, L. \& {Rosolowsky}, E. 2004, \apjl, 612, L29

\bibitem[{{Blitz} \& {Rosolowsky}(2006)}]{Blitz&Rosolowsky06}
---. 2006, \apj, 650, 933

\bibitem[{{Blondin} {et~al.}(1998){Blondin}, {Wright}, {Borkowski}, \&
  {Reynolds}}]{Blondin+98}
{Blondin}, J.~M., {Wright}, E.~B., {Borkowski}, K.~J., \& {Reynolds}, S.~P.
  1998, \apj, 500, 342

\bibitem[{{Bolatto} {et~al.}(2008){Bolatto}, {Leroy}, {Rosolowsky}, {Walter},
  \& {Blitz}}]{Bolattoetal08}
{Bolatto}, A.~D., {Leroy}, A.~K., {Rosolowsky}, E., {Walter}, F., \& {Blitz},
  L. 2008, \apj, 686, 948

\bibitem[{{Boulares} \& {Cox}(1990)}]{Boulares&Cox90}
{Boulares}, A. \& {Cox}, D.~P. 1990, \apj, 365, 544

\bibitem[{{Daddi} {et~al.}(2010){Daddi}, {Elbaz}, {Walter}, {Bournaud},
  {Salmi}, {Carilli}, {Dannerbauer}, {Dickinson}, {Monaco}, \&
  {Riechers}}]{Daddi+10}
{Daddi}, E., {Elbaz}, D., {Walter}, F., {Bournaud}, F., {Salmi}, F., {Carilli},
  C., {Dannerbauer}, H., {Dickinson}, M., {Monaco}, P., \& {Riechers}, D. 2010,
  \apjl, 714, L118

\bibitem[{{de Avillez} \& {Breitschwerdt}(2004)}]{deAvillez&Breitschwerdt04}
{de Avillez}, M.~A. \& {Breitschwerdt}, D. 2004, \aap, 425, 899

\bibitem[{{Dib} {et~al.}(2006){Dib}, {Bell}, \& {Burkert}}]{Dib+06}
{Dib}, S., {Bell}, E., \& {Burkert}, A. 2006, \apj, 638, 797

\bibitem[{{Dobbs}(2008)}]{Dobbs08}
{Dobbs}, C.~L. 2008, \mnras, 391, 844

\bibitem[{{Dobbs} {et~al.}(2011){Dobbs}, {Burkert}, \& {Pringle}}]{Dobbs+11b}
{Dobbs}, C.~L., {Burkert}, A., \& {Pringle}, J.~E. 2011, \mnras, 417, 1318

\bibitem[{{Dobbs} \& {Pringle}(2009)}]{Dobbs&Pringle09}
{Dobbs}, C.~L. \& {Pringle}, J.~E. 2009, \mnras, 396, 1579

\bibitem[{{Downes} \& {Solomon}(1998)}]{Downes&Solomon98}
{Downes}, D. \& {Solomon}, P.~M. 1998, \apj, 507, 615

\bibitem[{{Draine}(2011)}]{Drainebook11}
{Draine}, B.~T. 2011, {Physics of the Interstellar and Intergalactic Medium},
  ed. {Draine, B.~T.}

\bibitem[{{Evans} {et~al.}(2009){Evans}, {Dunham}, {J{\o}rgensen}, {Enoch},
  {Mer{\'{\i}}n}, {van Dishoeck}, {Alcal{\'a}}, {Myers}, {Stapelfeldt},
  {Huard}, {Allen}, {Harvey}, {van Kempen}, {Blake}, {Koerner}, {Mundy},
  {Padgett}, \& {Sargent}}]{Evans+09}
{Evans}, II, N.~J., {Dunham}, M.~M., {J{\o}rgensen}, J.~K., {Enoch}, M.~L.,
  {Mer{\'{\i}}n}, B., {van Dishoeck}, E.~F., {Alcal{\'a}}, J.~M., {Myers},
  P.~C., {Stapelfeldt}, K.~R., {Huard}, T.~L., {Allen}, L.~E., {Harvey}, P.~M.,
  {van Kempen}, T., {Blake}, G.~A., {Koerner}, D.~W., {Mundy}, L.~G.,
  {Padgett}, D.~L., \& {Sargent}, A.~I. 2009, \apjs, 181, 321

\bibitem[{{Gammie} {et~al.}(2003){Gammie}, {Lin}, {Stone}, \&
  {Ostriker}}]{Gammieetal03}
{Gammie}, C.~F., {Lin}, Y.-T., {Stone}, J.~M., \& {Ostriker}, E.~C. 2003, \apj,
  592, 203

\bibitem[{{Genzel} {et~al.}(2011){Genzel}, {Newman}, {Jones}, {F{\"o}rster
  Schreiber}, {Shapiro}, {Genel}, {Lilly}, {Renzini}, {Tacconi}, {Bouch{\'e}},
  {Burkert}, {Cresci}, {Buschkamp}, {Carollo}, {Ceverino}, {Davies}, {Dekel},
  {Eisenhauer}, {Hicks}, {Kurk}, {Lutz}, {Mancini}, {Naab}, {Peng},
  {Sternberg}, {Vergani}, \& {Zamorani}}]{Genzel+11}
{Genzel}, R., {Newman}, S., {Jones}, T., {F{\"o}rster Schreiber}, N.~M.,
  {Shapiro}, K., {Genel}, S., {Lilly}, S.~J., {Renzini}, A., {Tacconi}, L.~J.,
  {Bouch{\'e}}, N., {Burkert}, A., {Cresci}, G., {Buschkamp}, P., {Carollo},
  C.~M., {Ceverino}, D., {Davies}, R., {Dekel}, A., {Eisenhauer}, F., {Hicks},
  E., {Kurk}, J., {Lutz}, D., {Mancini}, C., {Naab}, T., {Peng}, Y.,
  {Sternberg}, A., {Vergani}, D., \& {Zamorani}, G. 2011, \apj, 733, 101

\bibitem[{{Genzel} {et~al.}(2010){Genzel}, {Tacconi}, {Gracia-Carpio},
  {Sternberg}, {Cooper}, {Shapiro}, {Bolatto}, {Bouch{\'e}}, {Bournaud},
  {Burkert}, {Combes}, {Comerford}, {Cox}, {Davis}, {Schreiber},
  {Garcia-Burillo}, {Lutz}, {Naab}, {Neri}, {Omont}, {Shapley}, \&
  {Weiner}}]{Genzel+10}
{Genzel}, R., {Tacconi}, L.~J., {Gracia-Carpio}, J., {Sternberg}, A., {Cooper},
  M.~C., {Shapiro}, K., {Bolatto}, A., {Bouch{\'e}}, N., {Bournaud}, F.,
  {Burkert}, A., {Combes}, F., {Comerford}, J., {Cox}, P., {Davis}, M.,
  {Schreiber}, N.~M.~F., {Garcia-Burillo}, S., {Lutz}, D., {Naab}, T., {Neri},
  R., {Omont}, A., {Shapley}, A., \& {Weiner}, B. 2010, \mnras, 407, 2091

\bibitem[{{Glover} \& {Mac Low}(2011)}]{Glover&MacLow11}
{Glover}, S.~C.~O. \& {Mac Low}, M. 2011, \mnras, 412, 337

\bibitem[{{Hennebelle} \& {Chabrier}(2011)}]{Hennebelle&Chabrier11}
{Hennebelle}, P. \& {Chabrier}, G. 2011, \apjl, 743, L29

\bibitem[{{Hill} {et~al.}(2012){Hill}, {Joung}, {Mac Low}, {Benjamin},
  {Haffner}, {Klingenberg}, \& {Waagan}}]{Hill+12}
{Hill}, A.~S., {Joung}, M.~R., {Mac Low}, M.-M., {Benjamin}, R.~A., {Haffner},
  L.~M., {Klingenberg}, C., \& {Waagan}, K. 2012, ArXiv e-prints

\bibitem[{{Hopkins} {et~al.}(2011){Hopkins}, {Quataert}, \&
  {Murray}}]{Hopkins+11}
{Hopkins}, P.~F., {Quataert}, E., \& {Murray}, N. 2011, \mnras, 417, 950

\bibitem[{{Joung} {et~al.}(2009){Joung}, {Mac Low}, \& {Bryan}}]{Joung+09}
{Joung}, M.~R., {Mac Low}, M.-M., \& {Bryan}, G.~L. 2009, \apj, 704, 137

\bibitem[{{Kennicutt}(1989)}]{Kennicutt89}
{Kennicutt}, Jr., R.~C. 1989, \apj, 344, 685

\bibitem[{{Kennicutt}(1998)}]{Kennicutt98}
---. 1998, \apj, 498, 541

\bibitem[{{Kim} {et~al.}(2011){Kim}, {Kim}, \& {Ostriker}}]{Kim+11}
{Kim}, C.-G., {Kim}, W.-T., \& {Ostriker}, E.~C. 2011, \apj, 743, 25 (KKO11)

\bibitem[{{Klessen}(2000)}]{Klessen00}
{Klessen}, R.~S. 2000, \apj, 535, 869

\bibitem[{{Koyama} \& {Ostriker}(2009{\natexlab{a}})}]{KoyamaOstriker09a}
{Koyama}, H. \& {Ostriker}, E.~C. 2009{\natexlab{a}}, \apj, 693, 1316

\bibitem[{{Koyama} \& {Ostriker}(2009{\natexlab{b}})}]{Koyama&Ostriker09b}
---. 2009{\natexlab{b}}, \apj, 693, 1346

\bibitem[{{Kroupa}(2001)}]{Kroupa01}
{Kroupa}, P. 2001, \mnras, 322, 231

\bibitem[{{Krumholz} {et~al.}(2012){Krumholz}, {Dekel}, \&
  {McKee}}]{Krumholz+12}
{Krumholz}, M.~R., {Dekel}, A., \& {McKee}, C.~F. 2012, \apj, 745, 69

\bibitem[{{Krumholz} \& {McKee}(2005)}]{Krumholz&McKee05}
{Krumholz}, M.~R. \& {McKee}, C.~F. 2005, \apj, 630, 250

\bibitem[{{Krumholz} {et~al.}(2009){Krumholz}, {McKee}, \&
  {Tumlinson}}]{Krumholz+09}
{Krumholz}, M.~R., {McKee}, C.~F., \& {Tumlinson}, J. 2009, \apj, 699, 850

\bibitem[{{Krumholz} \& {Tan}(2007)}]{Krumholz&Tan07}
{Krumholz}, M.~R. \& {Tan}, J.~C. 2007, \apj, 654, 304

\bibitem[{{Larson}(1981)}]{Larson81}
{Larson}, R.~B. 1981, \mnras, 194, 809

\bibitem[{{Li} {et~al.}(2005){Li}, {Mac Low}, \& {Klessen}}]{Lietal05}
{Li}, Y., {Mac Low}, M.-M., \& {Klessen}, R.~S. 2005, \apjl, 620, L19

\bibitem[{{Li} {et~al.}(2006){Li}, {Mac Low}, \& {Klessen}}]{Lietal06}
---. 2006, \apj, 639, 879

\bibitem[{{Mac Low} \& {Klessen}(2004)}]{MacLow&Klessen04}
{Mac Low}, M. \& {Klessen}, R.~S. 2004, Reviews of Modern Physics, 76, 125

\bibitem[{{McKee} \& {Ostriker}(2007)}]{McKee&Ostriker07}
{McKee}, C.~F. \& {Ostriker}, E.~C. 2007, \araa, 45, 565

\bibitem[{{McKee} \& {Ostriker}(1977)}]{McKee&Ostriker77}
{McKee}, C.~F. \& {Ostriker}, J.~P. 1977, \apj, 218, 148

\bibitem[{{Narayanan} {et~al.}(2011){Narayanan}, {Krumholz}, {Ostriker}, \&
  {Hernquist}}]{Narayanan+11a}
{Narayanan}, D., {Krumholz}, M., {Ostriker}, E.~C., \& {Hernquist}, L. 2011,
  \mnras, 418, 664

\bibitem[{{Narayanan} {et~al.}(2012){Narayanan}, {Krumholz}, {Ostriker}, \&
  {Hernquist}}]{Narayanan+12}
{Narayanan}, D., {Krumholz}, M.~R., {Ostriker}, E.~C., \& {Hernquist}, L. 2012,
  \mnras, 421, 3127

\bibitem[{{Norman} \& {Ferrara}(1996)}]{Norman&Ferrara96}
{Norman}, C.~A. \& {Ferrara}, A. 1996, \apj, 467, 280

\bibitem[{{Norman} \& {Ikeuchi}(1989)}]{Norman&Ikeuchi89}
{Norman}, C.~A. \& {Ikeuchi}, S. 1989, \apj, 345, 372

\bibitem[{{Oka} {et~al.}(1998){Oka}, {Hasegawa}, {Hayashi}, {Handa}, \&
  {Sakamoto}}]{Oka+98}
{Oka}, T., {Hasegawa}, T., {Hayashi}, M., {Handa}, T., \& {Sakamoto}, S. 1998,
  \apj, 493, 730

\bibitem[{{Oka} {et~al.}(2001){Oka}, {Hasegawa}, {Sato}, {Tsuboi}, {Miyazaki},
  \& {Sugimoto}}]{Oka+01}
{Oka}, T., {Hasegawa}, T., {Sato}, F., {Tsuboi}, M., {Miyazaki}, A., \&
  {Sugimoto}, M. 2001, \apj, 562, 348

\bibitem[{{Ostriker} {et~al.}(1999){Ostriker}, {Gammie}, \&
  {Stone}}]{Ostriker+99}
{Ostriker}, E.~C., {Gammie}, C.~F., \& {Stone}, J.~M. 1999, \apj, 513, 259

\bibitem[{{Ostriker} {et~al.}(2010){Ostriker}, {McKee}, \&
  {Leroy}}]{Ostriker+10}
{Ostriker}, E.~C., {McKee}, C.~F., \& {Leroy}, A.~K. 2010, \apj, 721, 975 (OML10)

\bibitem[{{Ostriker} \& {Shetty}(2011)}]{Ostriker&Shetty11}
{Ostriker}, E.~C. \& {Shetty}, R. 2011, \apj, 731, 41 (Paper I)

\bibitem[{{Ostriker} {et~al.}(2001){Ostriker}, {Stone}, \&
  {Gammie}}]{Ostriker+01}
{Ostriker}, E.~C., {Stone}, J.~M., \& {Gammie}, C.~F. 2001, \apj, 546, 980

\bibitem[{{Padoan} \& {Nordlund}(2011)}]{Padoan&Nordlund11b}
{Padoan}, P. \& {Nordlund}, {\AA}. 2011, \apj, 730, 40

\bibitem[{{Pichardo} {et~al.}(2000){Pichardo}, {V{\'a}zquez-Semadeni}, {Gazol},
  {Passot}, \& {Ballesteros-Paredes}}]{Pichardo+00}
{Pichardo}, B., {V{\'a}zquez-Semadeni}, E., {Gazol}, A., {Passot}, T., \&
  {Ballesteros-Paredes}, J. 2000, \apj, 532, 353

\bibitem[{{Robertson} \& {Kravtsov}(2008)}]{Robertson&Kravtsov08}
{Robertson}, B.~E. \& {Kravtsov}, A.~V. 2008, \apj, 680, 1083

\bibitem[{{Schmidt}(1959)}]{Schmidt59}
{Schmidt}, M. 1959, \apj, 129, 243

\bibitem[{{Schruba} {et~al.}(2011){Schruba}, {Leroy}, {Walter}, {Bigiel},
  {Brinks}, {de Blok}, {Dumas}, {Kramer}, {Rosolowsky}, {Sandstrom},
  {Schuster}, {Usero}, {Weiss}, \& {Wiesemeyer}}]{Schruba+11}
{Schruba}, A., {Leroy}, A.~K., {Walter}, F., {Bigiel}, F., {Brinks}, E., {de
  Blok}, W.~J.~G., {Dumas}, G., {Kramer}, C., {Rosolowsky}, E., {Sandstrom},
  K., {Schuster}, K., {Usero}, A., {Weiss}, A., \& {Wiesemeyer}, H. 2011, \aj,
  142, 37

\bibitem[{{Sheth} {et~al.}(2008){Sheth}, {Vogel}, {Wilson}, \&
  {Dame}}]{Sheth+08}
{Sheth}, K., {Vogel}, S.~N., {Wilson}, C.~D., \& {Dame}, T.~M. 2008, \apj, 675,
  330

\bibitem[{{Shetty} {et~al.}(2012){Shetty}, {Beaumont}, {Burton}, {Kelly}, \&
  {Klessen}}]{Shetty+12}
{Shetty}, R., {Beaumont}, C.~N., {Burton}, M.~G., {Kelly}, B.~C., \& {Klessen},
  R.~S. 2012, MNRAS submitted

\bibitem[{{Shetty} {et~al.}(2010){Shetty}, {Collins}, {Kauffmann}, {Goodman},
  {Rosolowsky}, \& {Norman}}]{Shetty+10}
{Shetty}, R., {Collins}, D.~C., {Kauffmann}, J., {Goodman}, A.~A.,
  {Rosolowsky}, E.~W., \& {Norman}, M.~L. 2010, \apj, 712, 1049

\bibitem[{{Shetty} {et~al.}(2011{\natexlab{a}}){Shetty}, {Glover}, {Dullemond},
  \& {Klessen}}]{Shetty+11a}
{Shetty}, R., {Glover}, S.~C., {Dullemond}, C.~P., \& {Klessen}, R.~S.
  2011{\natexlab{a}}, \mnras, 412, 1686

\bibitem[{{Shetty} {et~al.}(2011{\natexlab{b}}){Shetty}, {Glover}, {Dullemond},
  {Ostriker}, {Harris}, \& {Klessen}}]{Shetty+11b}
{Shetty}, R., {Glover}, S.~C., {Dullemond}, C.~P., {Ostriker}, E.~C., {Harris},
  A.~I., \& {Klessen}, R.~S. 2011{\natexlab{b}}, \mnras, 415, 3253

\bibitem[{{Shetty} \& {Ostriker}(2006)}]{Shetty&Ostriker06}
{Shetty}, R. \& {Ostriker}, E.~C. 2006, \apj, 647, 997

\bibitem[{{Shetty} \& {Ostriker}(2008)}]{Shetty&Ostriker08}
---. 2008, \apj, 684, 978

\bibitem[{{Solomon} {et~al.}(1997){Solomon}, {Downes}, {Radford}, \&
  {Barrett}}]{Solomonetal97}
{Solomon}, P.~M., {Downes}, D., {Radford}, S.~J.~E., \& {Barrett}, J.~W. 1997,
  \apj, 478, 144

\bibitem[{{Solomon} {et~al.}(1987){Solomon}, {Rivolo}, {Barrett}, \&
  {Yahil}}]{Solomonetal87}
{Solomon}, P.~M., {Rivolo}, A.~R., {Barrett}, J., \& {Yahil}, A. 1987, \apj,
  319, 730

\bibitem[{{Solomon} \& {Vanden Bout}(2005)}]{SolomonVandenBout05}
{Solomon}, P.~M. \& {Vanden Bout}, P.~A. 2005, \araa, 43, 677

\bibitem[{{Stone} \& {Gardiner}(2009)}]{Stone&Gardiner09}
{Stone}, J.~M. \& {Gardiner}, T. 2009, New Astronomy, 14, 139

\bibitem[{{Stone} {et~al.}(2008){Stone}, {Gardiner}, {Teuben}, {Hawley}, \&
  {Simon}}]{Stone+08}
{Stone}, J.~M., {Gardiner}, T.~A., {Teuben}, P., {Hawley}, J.~F., \& {Simon},
  J.~B. 2008, \apjs, 178, 137

\bibitem[{{Tasker}(2011)}]{Tasker11}
{Tasker}, E.~J. 2011, \apj, 730, 11

\bibitem[{{Tasker} \& {Bryan}(2006)}]{Tasker&Bryan06}
{Tasker}, E.~J. \& {Bryan}, G.~L. 2006, \apj, 641, 878

\bibitem[{{Tasker} \& {Bryan}(2008)}]{Tasker&Bryan08}
---. 2008, \apj, 673, 810

\bibitem[{{Tasker} \& {Tan}(2009)}]{Tasker&Tan09}
{Tasker}, E.~J. \& {Tan}, J.~C. 2009, \apj, 700, 358

\bibitem[{{Thompson} {et~al.}(2005){Thompson}, {Quataert}, \&
  {Murray}}]{Thompson+05}
{Thompson}, T.~A., {Quataert}, E., \& {Murray}, N. 2005, \apj, 630, 167

\bibitem[{{Toomre}(1964)}]{Toomre64}
{Toomre}, A. 1964, \apj, 139, 1217

\bibitem[{{Truelove} {et~al.}(1997){Truelove}, {Klein}, {McKee}, {Holliman},
  {Howell}, \& {Greenough}}]{Trueloveetal97}
{Truelove}, J.~K., {Klein}, R.~I., {McKee}, C.~F., {Holliman}, II, J.~H.,
  {Howell}, L.~H., \& {Greenough}, J.~A. 1997, \apjl, 489, L179+

\bibitem[{{Vazquez-Semadeni}(1994)}]{Vazquez-Semadeni94}
{Vazquez-Semadeni}, E. 1994, \apj, 423, 681

\bibitem[{{Walters} \& {Cox}(2001)}]{Walters&Cox01}
{Walters}, M.~A. \& {Cox}, D.~P. 2001, \apj, 549, 353

\bibitem[{{Wong} \& {Blitz}(2002)}]{Wong&Blitz02}
{Wong}, T. \& {Blitz}, L. 2002, \apj, 569, 157

\bibitem[{{Yusef-Zadeh} {et~al.}(2009){Yusef-Zadeh}, {Hewitt}, {Arendt},
  {Whitney}, {Rieke}, {Wardle}, {Hinz}, {Stolovy}, {Lang}, {Burton}, \&
  {Ramirez}}]{Yusef-Zadeh+09}
{Yusef-Zadeh}, F., {Hewitt}, J.~W., {Arendt}, R.~G., {Whitney}, B., {Rieke},
  G., {Wardle}, M., {Hinz}, J.~L., {Stolovy}, S., {Lang}, C.~C., {Burton},
  M.~G., \& {Ramirez}, S. 2009, \apj, 702, 178

\end{thebibliography}


\end{document}